\newcommand{\ha}{H$\alpha$}
\newcommand{\hb}{H$\beta$}
\newcommand{\oii}{[O\,{\sc ii}]}
\newcommand{\oiii}{[O\,{\sc iii}]}
\newcommand{\nii}{[N\,{\sc ii}]}
\newcommand{\sii}{[S\,{\sc ii}]}

\documentclass[iop,appendixfloats,numberedappendix]{emulateapj}

\shorttitle{The MOSDEF Survey}
\shortauthors{Kriek et al.}
\usepackage{natbib}
\slugcomment{Accepted for publication in ApJS}

\begin{document}
  
\title{The MOSFIRE Deep Evolution Field (MOSDEF) Survey: \\ Rest-frame Optical Spectroscopy for $\sim$1500 $H$-Selected Galaxies at $1.37\le\lowercase{z}\le3.8$}

\author{Mariska Kriek\altaffilmark{1}, Alice E. Shapley\altaffilmark{2}, Naveen A. Reddy\altaffilmark{3,4}, Brian Siana\altaffilmark{3}, Alison L. Coil\altaffilmark{5}, Bahram Mobasher\altaffilmark{3}, William R. Freeman\altaffilmark{3}, Laura de Groot\altaffilmark{3}, Sedona H. Price\altaffilmark{1}, Ryan Sanders\altaffilmark{2}, Irene Shivaei\altaffilmark{3}, Gabriel B. Brammer\altaffilmark{6}, Ivelina G. Momcheva\altaffilmark{7}, Rosalind E. Skelton\altaffilmark{8}, Pieter G. van Dokkum\altaffilmark{7}, Katherine E. Whitaker\altaffilmark{9}, James Aird\altaffilmark{10}, Mojegan Azadi\altaffilmark{5}, Marc Kassis\altaffilmark{11}, James S. Bullock\altaffilmark{12}, Charlie Conroy\altaffilmark{13,14}, Romeel Dav\'e\altaffilmark{15}, Du{\v s}an Kere{\v s}\altaffilmark{5}, and Mark Krumholz\altaffilmark{14}}

\altaffiltext{1}{Astronomy Department, University of California, Berkeley, CA 94720, USA}

\altaffiltext{2}{Department of Physics \& Astronomy, University of California, Los Angeles, CA 90095, USA}

\altaffiltext{3}{Department of Physics \& Astronomy, University of California, Riverside, CA 92521, USA}

\altaffiltext{4}{Alfred P. Sloan Research Fellow}

\altaffiltext{5}{Center for Astrophysics and Space Sciences, University of California, San Diego, La Jolla, CA 92093, USA}

\altaffiltext{6}{Space Telescope Science Institute, Baltimore, MD 21218, USA}

\altaffiltext{7}{Department of Astronomy, Yale University, New Haven, CT 06511, USA}

\altaffiltext{8}{Department of Astronomy, University of Cape Town, Private Bag X3, Rondebosch 7701, South Africa}

\altaffiltext{9}{Astrophysics Science Division, Goddard Space Center, Greenbelt, MD 20771, USA}

\altaffiltext{10}{Institute of Astronomy, University of Cambridge, Madingley Road, Cambridge CB3 0HA, UK}

\altaffiltext{11}{W.M. Keck Observatory, Kamuela, HI 96743-8431, USA}

\altaffiltext{12}{Center for Cosmology, Department of Physics and Astronomy, University of California, Irvine, CA 92697, USA}

\altaffiltext{13}{Harvard-Smithsonian Center for Astrophysics, 60 Garden St., Cambridge, MA, USA}

\altaffiltext{14}{Department of Astronomy \& Astrophysics, University of California, Santa Cruz, CA, USA}

\altaffiltext{15}{University of the Western Cape, Bellville, Cape Town 7535, South Africa}

\begin{abstract}
In this paper we present the MOSFIRE Deep Evolution Field (MOSDEF)
survey. The MOSDEF survey aims to obtain moderate-resolution
($R=3000-3650$) rest-frame optical spectra ($\sim3700-7000$~\AA) for
$\sim$1500 galaxies at $1.37\le\lowercase{z}\le3.80$ in three
well-studied CANDELS fields: AEGIS, COSMOS, and GOODS-N. Targets are
selected in three redshift intervals: $1.37\le z\le 1.70$, $2.09\le
z\le2.61$, and $2.95\le z\le 3.80$, down to fixed $H_{\rm AB}$~(F160W)
magnitudes of 24.0, 24.5 and 25.0, respectively, using the photometric
and spectroscopic catalogs from the 3D-HST survey. We target both
strong nebular emission lines (e.g., \oii\,$\lambda\lambda$3727,3730,
H$\beta$, \oiii\,$\lambda\lambda4960,5008$, H$\alpha$,
\nii\,$\lambda\lambda6550,6585$, and \sii\,$\lambda\lambda6718,6733$)
and stellar continuum and absorption features (e.g., Balmer lines,
Ca-{\sc ii}~H and K, Mgb, 4000\,\AA~break). Here we present an
overview of our survey, the observational strategy, the data reduction
and analysis, and the sample characteristics based on spectra obtained
during the first 24 nights. To date, we have completed 21 masks,
obtaining spectra for 591 galaxies. For $\sim$80\% of the targets we
 derive a robust redshift from either emission or absorption
lines. In addition, we confirm 55 additional galaxies, which were
serendipitously detected. The MOSDEF galaxy sample includes unobscured
star-forming, dusty star-forming, and quiescent galaxies and spans a
wide range in stellar mass ($\sim\,10^{9}-10^{11.5}~M_{\odot}$) and
star formation rate ($\sim\,10^{0}-10^{3}~M_{\odot}\rm~yr^{-1}$). The
spectroscopically confirmed sample is roughly representative of an
H-band limited galaxy sample at these redshifts. With its large sample
size, broad diversity in galaxy properties, and wealth of available
ancillary data, MOSDEF will transform our understanding of the
stellar, gaseous, metal, dust, and black hole content of galaxies
during the time when the universe was most active. 
\end{abstract} 


\section{INTRODUCTION}\label{sec:int}

Understanding the formation and evolution of galaxies remains one of the 
greatest challenges of modern astronomy. Key outstanding questions include: 
What are the physical processes driving star formation 
in individual galaxies? How do galaxies exchange gas and heavy elements 
with the intergalactic medium? How are stellar mass and structure 
assembled in galaxies (in situ star formation versus mergers)?  What 
is the nature of the co-evolution of black holes and stellar 
populations?

Addressing these questions requires observations of galaxy populations
across cosmic time. The Sloan Digital Sky
Survey \citep[SDSS;][]{DYork2000} and the 2dF Galaxy Redshift
Survey \citep{MColless2001} provide a detailed description of
the local galaxy population, with imaging and spectra of more than
$10^{6}$ galaxies. These data  quantify the distributions in galaxy
luminosity, color, stellar, dynamical and  black hole mass, structural
properties, gas content, metallicity, and  environment, as well as the
strong correlations among these parameters. Such results provide an
endpoint for our description  of galaxy evolution. 

In order to understand the full story from beginning to end, however,
we require observations probing earlier cosmic epochs. Several
spectroscopic surveys \citep[e.g., DEEP2, VVDS, zCOSMOS,
  PRIMUS;][]{JNewman2013,OLeFevre2005,SLilly2007,ACoil2011} have
probed the properties of galaxy populations to $z\sim 1$ with sample
sizes of $\sim 10^4-10^5$ objects, describing the evolution in the
luminosities, colors, stellar masses, sizes, and environments of both
star-forming and quiescent galaxies over the past $\sim 8$~Gyr. 

The next frontier for comprehensive galaxy surveys is the epoch at
$1.5 \lesssim z \lesssim 3.5$, the peak of both star formation and black hole
accretion activity in the
universe \cite[e.g.,][]{AHopkins2006,NReddy2008}. Several
qualitative imprints of the local galaxy patterns have already been
observed at these earlier times, including the bimodal distribution of
galaxy
colors \citep[e.g.,][]{PCassata2008,MKriek2008b,RWilliams2009,KWhitaker2011},
the strong clustering of red
galaxies \citep[e.g.,][]{RQuadri2008}, the mass-metallicity
relation \citep[e.g.,][]{DErb2006a}, and the correlation
between stellar population properties and structural
parameters \citep[e.g.,][]{RWilliams2010,SWuyts2011b}. However,
there are also striking differences, such as the large diversity among
massive
galaxies \citep[e.g.,][]{MKriek2009a,PvanDokkum2011,AMuzzin2013b}
and the absence of cold, quiescent disk
galaxies \citep[e.g.,][]{NForsterSchreiber2009,DLaw2009}. 

Although great strides have been made in the past several years to
survey galaxies at this key epoch \citep[see][for a recent
  review]{AShapley2011}, most studies are based on multi-wavelength
photometric data alone, with little or no spectroscopic information
\citep[e.g.,][]{SWuyts2011a}. Rest-frame UV spectra have been measured
for $\sim 3000$ galaxies at $1.5\le z\le 3.5$
\citep{CSteidel2003,CSteidel2004}, yet the sample of such objects is
biased towards relatively  blue, star-forming galaxies, and these
spectra are primarily sensitive to interstellar and circumgalactic
medium (ISM/CGM) features tracing outflowing gas
\citep[e.g.,][]{AShapley2003}. Current surveys with the  {\it
  HST}/WFC3 near-IR grisms are yielding rest-frame optical spectra of
$\sim 10,000$ galaxies at $z>1$ \citep{GBrammer2012,HAtek2010}, yet
the low resolution ($R< 130$) and limited wavelength range ($\lambda
<1.6 \mu$m) prevent a robust characterization of emission and
absorption line ratios and widths over the range $1.5\lesssim
z\lesssim 3.5$. Finally, moderate-resolution rest-frame optical
spectroscopy has been obtained at $1.5\lesssim z\lesssim 3.5$ for (UV-selected)
star-forming galaxies
\citep[e.g.,][]{DErb2006a,FMannucci2009,NForsterSchreiber2009} and
stellar mass-limited samples of massive galaxies \citep[$M\geq
  10^{11}\,M_{\odot}$; e.g.,][]{MKriek2008a}. Yet the largest
homogeneous sample obtained until recently consisted of $\sim100$ galaxies, and
the wavelength coverage and depth was in most cases insufficient to
observe all strong spectral features.

Key requirements for a complete evolutionary census of the galaxy
population at $1.5\lesssim z \lesssim 3.5$ include: (1) rest-frame
optical spectroscopy covering all of the strongest emission and
absorption features between rest-frame 3700 and 6800~\AA, with
sufficient resolution to characterize the gaseous and stellar contents
of galaxies; (2) a large  ($N>10^3$) sample of objects, spanning the
full diversity of stellar populations and dust extinction over a large
dynamic range in stellar mass; and (3) multiple redshift bins to
enable evolutionary studies. These requirements can now be met with
the commissioning of the MOSFIRE spectrograph
\citep{IMcLean2010,IMcLean2012} on the Keck~I telescope. KMOS
\citep{RSharples2004} on the VLT and FMOS \citep{MKimura2010} on
Subaru are also providing near-IR spectra of large samples of distant
galaxies \citep[e.g.,][]{EWisnioski2015,JSilverman2014}.

MOSFIRE is a multi-object moderate resolution spectrograph operating
from $0.97$ to $2.45\; \mu$m, enabling the simultaneous spectroscopic
observation of $\sim$30 individual galaxies distributed over a
$6' \times 3'$ field of view. \cite{CSteidel2014} demonstrate
the power of MOSFIRE for studying nebular line emission using a sample
of 179 star-forming galaxies at $2<z<3$. In
addition, \cite{SBelli2014} illustrate MOSFIRE's potential
for continuum emission studies of distant galaxies, using a sample of
6 massive galaxies at $2<z<3$. To combine these two strengths and
assemble the first true statistical and magnitude-limited
spectroscopic galaxy sample at these redshifts, we are conducting the
MOSFIRE Deep Evolution Field (MOSDEF) survey. We plan to obtain
rest-frame optical spectra for $\sim$1500 galaxies in the range
$1.4\le z\le 3.8$. Together with existing multi-wavelength
data, MOSDEF enables measurements of the stellar, gaseous, metal,
dust, and black hole content of galaxies spanning a wider
dynamic range in physical properties than has ever been accessed
before with rest-frame optical spectroscopic surveys at these
redshifts. 

\begin{figure*}
  \begin{center}  
  \includegraphics[width=0.8\textwidth]{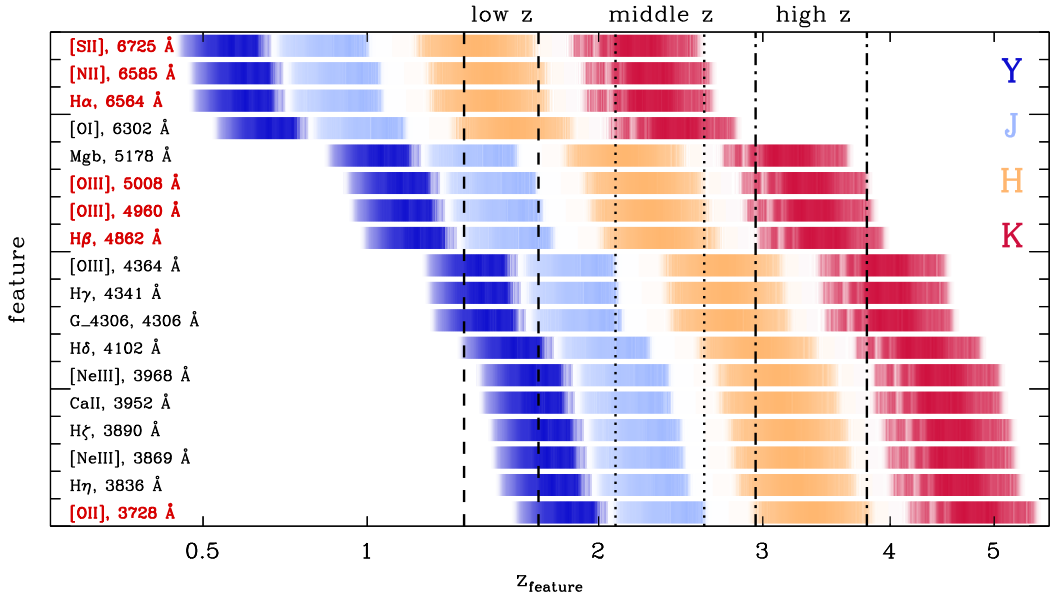}     

  \caption{MOSFIRE visibility of various rest-frame optical emission
    and absorption features as a function of redshift. Each row
    represents a different spectral feature, as indicated on the
    left. The primary emission line features are indicated in bold
    red. Each color represents a different filter, as indicated in the
    top right. The response curves are used for each filter and
    feature, and thus the brightness reflects the relative throughput
    at the corresponding redshift. The MOSDEF low ($1.37\le z\le
    1.70$), middle ($2.09\le z\le 2.61$) and high ($2.95\le z\le
    3.80$) redshift intervals are indicated by the dashed, dotted, and
    dashed-dotted vertical lines, respectively. This figure
    illustrates that the MOSDEF survey covers key emission features in
    each target redshift interval. \label{fig:features}}

\end{center}  
\end{figure*}

MOSDEF is being executed over 47 nights from December 2012 to
the spring of 2016. In this paper we present an overview of the
survey, the observational strategy, the data reduction and analysis,
and the sample characteristics based on data obtained over the first
24 nights of observing. The paper is organized as follows. In
Section~\ref{sec:obs} we present the MOSDEF survey design and
observing strategy, and an overview of the first observing runs. In
Section~\ref{sec:red} we discuss the two-dimensional (2D) data
processing, the noise properties, and the extraction of the
one-dimensional (1D) spectra. Section~\ref{sec:ana} describes the
spectral measurements, line and continuum sensitivities, the
spectroscopic success rate, the sample characteristics, and a
comparison to the parent magnitude-limited sample at the same
redshifts, from which targets are drawn. In Section~\ref{sec:sci} we
outline the MOSDEF science objectives, and finally, in
Section~\ref{sec:sum} we present a summary. 

Throughout this work we assume a $\Lambda$CDM cosmology with
$\Omega_{\rm m}$ = 0.3, $\Omega_{\rm \Lambda} = 0.7$,and $H_0 = 70$ km
s$^{-1}$ Mpc$^{-1}$. All magnitudes are given in the AB-magnitude
system \citep{joke1983}. The wavelengths of all emission and absorption
lines are given in vacuum. 

\section{OBSERVATIONS}\label{sec:obs}

\subsection{Survey Design}\label{sec:surdes}

In order to quantify galaxy evolution within the MOSDEF survey, we
target 3 redshift ranges, $1.37\le z\le 1.70$, $2.09\le z\le 2.61$,
$2.95\le z\le 3.80$. The targeted redshift regimes are selected such
that bright rest-frame optical emission lines fall within atmospheric
windows, as illustrated in Figure~\ref{fig:features}. A key aspect of
our survey strategy is that we cover multiple rest-frame optical
emission lines for each galaxy. Thus our strategy requires 2 or 3
filters per slit mask. For the $1.37\le z\le 1.70$ interval, we target
\hb\ and \oiii\ in the J-band and H$\alpha$, \nii, and \sii\ in the H
band. Within this window, we also target \oii\ in the Y-band for
$1.61\le z\le 1.70$. These same features appear in J, H, and K for the
$2.09\le z\le 2.61$ interval. For the $2.95\le z\le 3.80$ interval,
\oii\ falls in the H-band, and H$\beta$ and \oiii\ in the K-band. We
also target several continuum features and absorption lines in each of
our three redshift regimes, including the 4000\,\AA\ break, Ca\,{\sc
  ii} H and K, Mgb at 5178\,\AA, and Balmer absorption lines.

Emphasis is given to the middle redshift regime ($2.09\leq z\leq
2.61$), in which we plan to obtain a total sample of $\sim$750
galaxies. In the low and high redshift ranges, together, we aim to
target $\sim750$ galaxies as well, with the sample being roughly equally
split between the two intervals. As each redshift interval requires
different filter combinations, we use different masks for each
interval. Nonetheless, we include targets from the other redshift
intervals as fillers as space allows on each mask. The total planned
area is $\sim$600 square arcmin for the middle redshift regime, and
$\sim$300 square arcmin for the lower and higher redshift regimes. 

\begin{figure*}
  \begin{center}  
  \includegraphics[width=0.33\textwidth]{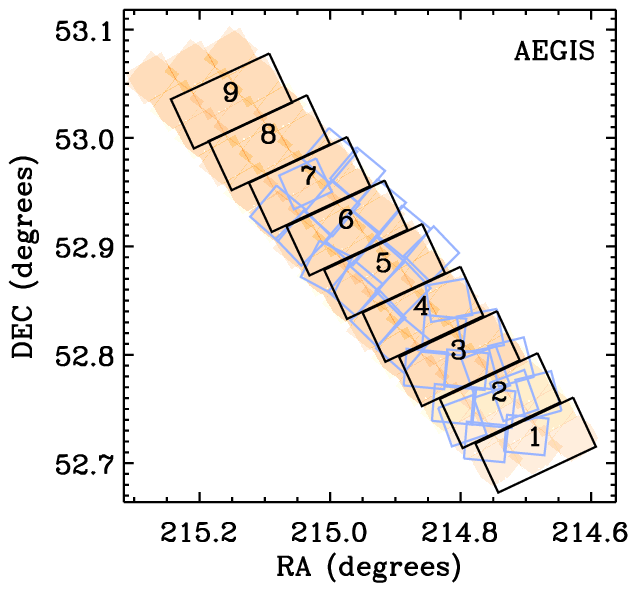}        
  \includegraphics[width=0.33\textwidth]{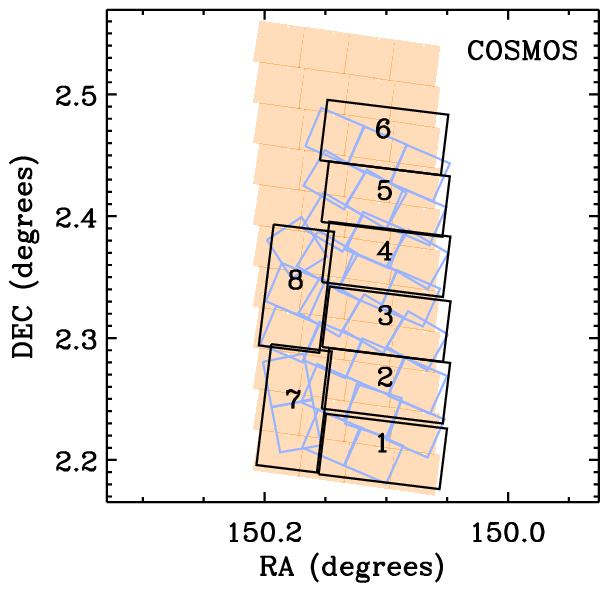}
  \includegraphics[width=0.33\textwidth]{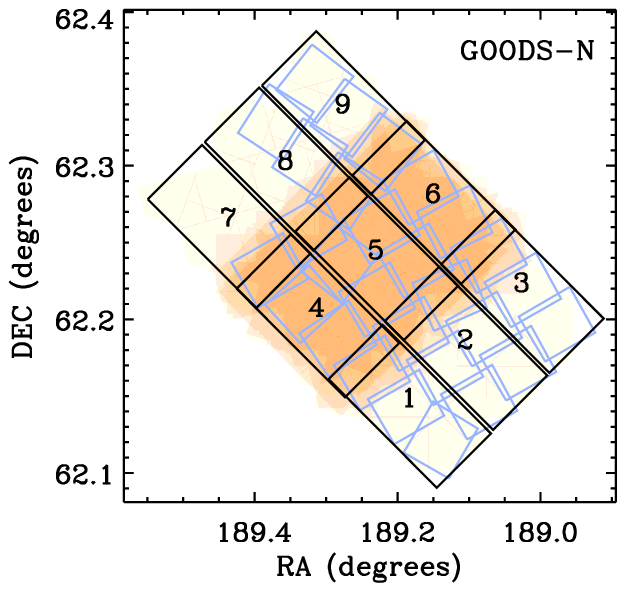}

  \caption{Footprints of MOSDEF observations in the primary target
    fields. From left to right we show the AEGIS, COSMOS, and GOODS-N
    fields. The MOSFIRE pointings and corresponding numbers are
    indicated in black. Each MOSFIRE pointing is $3'\times 6'$. The
    shaded orange regions represent the CANDELS {\it HST} WFC3/F160W
    exposure maps, and the open light blue boxes represent the 3D-HST
    WFC3/G141 grism pointings. \label{fig:fields}}

\end{center}  
\end{figure*}

The MOSDEF survey is primarily being executed in three well-studied
legacy fields with deep extensive multi-wavelength datasets: AEGIS
\citep{MDavis2007}, COSMOS \citep{NScoville2007}, and GOODS-N
\citep{MGiavalisco2004}. During the first observing season we also
observed one mask in UKIDSS-UDS \citep{ALawrence2007} and one mask in
GOODS-S \citep{MGiavalisco2004}, as our primary target fields were not
visible during the first half of the night. Within all fields we
target the regions that are covered by the CANDELS
\citep{NGrogin2011,AKoekemoer2011} and 3D-HST \citep{GBrammer2012}
surveys, as illustrated in Figure~\ref{fig:fields} for the three
primary fields. 

Targets are selected using photometric catalogs and grism spectra as
provided by the 3D-HST collaboration. These catalogs contain all
public photometric data and spectroscopic redshifts available for the
MOSDEF survey fields. A description of the photometric catalogs is
given in \cite{RSkelton2014}, and the grism spectra are described in
\cite{GBrammer2012}. The grism redshifts are derived by fitting the
grism spectra and multi-wavelength photometry simultaneously
\citep[][Momcheva et al. in
  prep]{GBrammer2012,GBrammer2013}. Additional spectroscopic redshifts
are included as well
\citep{NReddy2006,ABarger2008,ACoil2009,ACoil2011,MCooper2012,JNewman2013,JvandeSande2013}. When
no spectroscopic redshift is available, from either the grism
spectra or other spectroscopic campaigns, we use a photometric
redshift as derived using EAzY \citep{GBrammer2008}. 

Within each redshift interval we select by H-band (F160W)
magnitude. The magnitude limits are $H=24.0$, $H=24.5$, and $H=25.0$,
for the lower, middle and higher redshift intervals, respectively. For
these limits we obtain a roughly consistent stellar mass limit of
$\sim10^9\,M_\odot$ in each redshift interval. The 3D-HST catalogs
used for the selection are F125W+F140W+F160W-selected, and
\cite{RSkelton2014} show that the catalogs are 90\% complete at a
magnitude of $H=25$ for the shallow CANDELS data. GOODS-N is complete
to a fainter magnitude. The H-band covers the rest-frame optical
wavelength regime out to $z\sim3.8$, and thus we target a wide range
in galaxy spectral energy distributions (SEDs). However, as discussed
in Section~\ref{sec:samplechar}, this selection will slightly bias our
sample to unobscured star-forming galaxies with lower mass-to-light
ratios ($M/L_{\rm H}$) in the H-band. Nonetheless, within the
star-forming population, the rest-optical is less biased to recent
bursts of star formation, as changes in $M/L_{\rm H}$ are small
compared to changes in the equivalent widths of the nebular emission
lines or the UV and infrared continua \citep[e.g.,][]{ADominguez2014}.

To increase the success rate of our survey, we prioritize by magnitude
and redshift when selecting targets for spectroscopy. The initial
priority scales with H-band magnitude, such that brighter galaxies
have a higher priority. This criterion ensures that we obtain spectra
of the rare, massive galaxies and that our sample is not dominated by
the far more numerous galaxies near the flux limit. Within a given
magnitude bin we further prioritize by redshift. The highest priority
is given to galaxies with robust spectroscopic redshifts, from either
grism emission lines or other spectroscopic campaigns. Next, we
prioritize galaxies which have grism redshifts (without emission
lines) or photometric redshifts in the middle of the redshift
intervals: $1.42\le z\le1.65$, $2.20\le z\le 2.50$, and $3.05\le z
\le3.70$. The lowest priority is given to galaxies with photometric
redshifts at the edges of each interval. Within the lowest redshift
interval, we further prioritize galaxies at $1.61\le z\le 1.70$, for
which \oii\ is expected in the Y-band. Within a specific magnitude and
redshift bin, we upweight galaxies that host an AGN, identified either
by IRAC colors \citep{JDonley2012} or by a strong X-ray
counterpart. Details on the AGN selection are discussed in
\cite{ACoil2015}. Finally, for each target, we inspect the F160W
image, broadband SED, best-fit stellar population model, and grism
spectrum when available. In cases where the target looks unreliable
(e.g., very noisy photometry, mismatch between the photometry and the
best-fit stellar population model, mis-identified grism lines), it is
replaced by another target during mask design (see next section).  On
average we replaced 1-2 targets per mask. However, the exact number
depends on the photometric depth, which varies with field and targeted
redshift interval.

\subsection{Observing Strategy}
\label{sec:obsstr}

Our survey is being executed with MOSFIRE \citep{IMcLean2012} on the
Keck I telescope. MOSFIRE is a multi-object near-IR spectrograph with
an effective field of view of 3\arcmin\ by 6\arcmin. MOSFIRE has a
cryogenic configurable slit unit (CSU), consisting of 46 pairs of bars
of 7\farcs 1 length each. The bars can be configured in the horizontal
direction anywhere within the field, and can be combined in the
vertical direction with adjacent bars to make longer slits. We design
our masks using the
MAGMA\footnote{\href{http://www2.keck.hawaii.edu/inst/mosfire/magma.html}{http://www2.keck.hawaii.edu/inst/mosfire/magma.html}}
slitmask design software, and adopt a slit width of 0\farcs 7. This
slit width results in a spectral resolution of $R=3400,~3000,~3650$
and 3600 for Y, J, H, and K, respectively. The wavelength coverage
with a minimum of 5\% transmission is $0.962-1.135~\mu$m,
$1.142-1.365~\mu$m, $1.450-1.826~\mu$m, and $1.897-2.427~\mu$m, for Y,
J, H, and K, respectively. The actual wavelength coverage of the
spectra depends on the horizontal position of the slit in the CSU, and
thus it differs slightly among the different targets.

We use an ABA\arcmin B\arcmin\ ($+1\farcs 5, -1\farcs 2, +1\farcs 2,
-1\farcs 5$) dither pattern in order to increase the S/N of the final
spectra and to account for sky variations and detector defects
\citep[see][]{MKriek2008a}. However, during the first observing run we
also experimented with an ABBA dither pattern (see note to
Table~\ref{tab:summary}). The maximum offset for both of our dither
sequences is 3\arcsec. We set the dither space parameter in the MAGMA
software to 2\farcs 5, with the result that the center of each
target is at least 0\farcs 8 from the edge of each slit. While in
principle 46 objects can be observed at the same time, this small
dither space and the distribution of targets on the sky allow us to
observe on average 28 galaxies per mask, resulting in the assignment
of multiple pairs of bars for some slits. 

We adopt the individual exposure times recommended by the MOSFIRE
instrument team: 180 sec, 120 sec, 120 sec, and 180 sec, respectively,
for Y, J, H, and K. Using Fowler sampling with 16 readouts, these
integration times result in background-limited observations. The
corresponding gain and readout noise are 2.15 e-/cts and 5.8 e-,
respectively. The total nominal integration times are 1 hour per
filter for the $z\sim1.5$ masks, and 2 hours per filter for the
$z\sim2.3$ and $z\sim3.3$ masks.

We use a minimum of 5 alignment stars to acquire the slit masks with
an H-band magnitude between 18 and 21. However, stars with $H>20.5$
were too faint to be used for the alignment when conditions were
non-optimal in terms of seeing and transperancy. All but one of the
star boxes are replaced by science slits after alignment.  A slit is
configured on the remaining star in order to monitor the image
quality, throughput, and pointing accuracy. These slit stars are also
essential for the final flux calibration of the spectra (see
Section~\ref{sec:red}). For our pilot observing run in December 2012
we required an H-band magnitude $H<20.5$ for the slit stars. However,
we determined this limit to be too faint, and adjusted it to
$H\lesssim 20$ from 2013A onwards.

\begin{table*}[t!]
\centering
\caption{Mask Overview and Observations\smallskip}
\begin{tabular}{r | r r r | r  r  r  r | r  r  r  r |  r r r r | r}
\hline
Mask\footnote{The mask names include the targeted field, with ae: AEGIS;
co: COSMOS, gn: GOODS-N, gs: GOODS-S, and ud: UDS, the targeted redshift with 
1: $1.38\le z\le 1.70$, 2: $2.09\le z\le 2.61$, and 3: $2.95\le z\le 3.80$, and the
pointing as presented in Figure~\ref{fig:fields}.}
& \multicolumn{3}{c|}{Mask parameters}
& \multicolumn{4}{c|}{Integration times\footnote{The integration time only includes frames
that have been used in the reduction.}}
& \multicolumn{4}{c|}{FWHM seeing\footnote{The full width at half maximum (FWHM) of the seeing, measured from the slit star profile in
the reduced spectra.}} & \multicolumn{4}{c|}{3$\sigma$ depth\footnote{Measured from the noise frame of the reduced data, using optimal extraction and the slit star profile. Thus the quoted depths are for a faint point source only.}}& $N_{\rm
g}$\footnote{Number of targeted galaxies per mask. This number does not include
serendipitous detections or slit stars.} \\
& \multicolumn{1}{c}{RA} & \multicolumn{1}{c}{DEC}
& \multicolumn{1}{c|}{PA} & \multicolumn{4}{c|}{(min)}
&  \multicolumn{4}{c|}{(\arcsec)} & \multicolumn{4}{c|}{(AB mag)} & \\
& \multicolumn{1}{r}{(hh:mm:ss)} & \multicolumn{1}{r}{(dd:mm:ss)}
& \multicolumn{1}{c|}{($\deg$)} & Y & J & H & K & Y & J & H & K & Y &
J & H & K & \\
\hline\hline
$z\sim1.5$ &          &                &        &       &       &       &       &      &      &      &      &      &      &      &      &  \\          
 ae1\_01 &  14:18:45.13 &  52:43: 4.62 &  290.05 &    59.6 &    59.6 &    59.6 &       - &    0.62 &    0.52 &    0.54 &       - &    21.8 &    21.8 &    21.5 &       - &    28 \\
 ae1\_05 &  14:19:39.72 &  52:52:37.07 &  290.38 &    71.6 &   127.2 &    95.4 &       - &    0.97 &    0.80 &    0.77 &       - &    21.5 &    21.9 &    21.2 &       - &    28 \\
 co1\_03 &  10: 0:24.43 &   2:18:38.86 &   78.05 &       - &   153.1 &   157.1 &       - &       - &    0.79 &    0.90 &       - &       - &    21.7 &    21.4 &       - &    29 \\
 co1\_05 &  10: 0:24.84 &   2:24:57.96 &   79.04 &       - &    87.5 &    55.7 &       - &       - &    0.87 &    0.83 &       - &       - &    21.4 &    21.3 &       - &    26 \\
 gn1\_04 &  12:37:10.32 &  62:11:58.99 &   49.29 &    59.6 &    59.6 &    59.6 &       - &    0.61 &    0.55 &    0.54 &       - &    21.8 &    21.7 &    21.6 &       - &    27 \\
 ud1\_01 &   2:17:36.71 &  -5:12:13.90 &   37.10 &    62.6 &    67.6 &    59.6 &       - &    0.99 &    0.52 &    0.65 &       - &    20.7 &    21.4 &    21.1 &       - &    33 \\

\hline
$z\sim2.3$ &          &                &        &       &       &       &       &      &      &      &      &      &      &      &      & \\          
 ae2\_03 &  14:19:13.54 &  52:47:51.46 &  297.64 &       - &   119.3 &   119.3 &   152.1 &       - &    0.80 &    0.55 &    0.67 &       - &    21.4 &    21.7 &    21.0 &    27 \\
 ae2\_04 &  14:19:25.40 &  52:50:14.18 &  295.33 &       - &   119.3 &   119.3 &   119.3 &       - &    0.71 &    0.63 &    0.72 &       - &    21.6 &    22.0 &    21.1 &    29 \\
 ae2\_05 &  14:19:36.52 &  52:52:29.47 &  291.04 &       - &   123.3 &   119.3 &   119.3 &       - &    0.77 &    0.76 &    0.76 &       - &    21.9 &    22.1 &    21.1 &    29 \\
 co2\_01 &  10: 0:24.77 &   2:12:24.66 &   78.38 &       - &   119.3 &   119.3 &   244.6 &       - &    0.61 &    0.49 &    0.63 &       - &    21.9 &    22.2 &    21.3 &    31 \\
 co2\_03 &  10: 0:24.13 &   2:18:28.36 &   90.30 &       - &   113.3 &   119.3 &   104.4 &       - &    0.64 &    0.58 &    0.63 &       - &    21.9 &    22.4 &    21.2 &    25 \\
 co2\_04 &  10: 0:24.03 &   2:21:53.56 &   79.37 &       - &   115.3 &   119.3 &   119.3 &       - &    0.54 &    0.56 &    0.49 &       - &    22.0 &    22.1 &    21.6 &    27 \\
 gn2\_04 &  12:37:15.91 &  62:12:10.99 &   47.31 &       - &   119.3 &   119.3 &   119.3 &       - &    0.63 &    0.66 &    0.50 &       - &    22.1 &    21.8 &    21.4 &    27 \\
 gn2\_05 &  12:36:59.42 &  62:14:31.79 &   43.02 &       - &   119.3 &   119.3 &   122.3 &       - &    0.73 &    0.67 &    0.48 &       - &    21.8 &    21.8 &    21.4 &    28 \\
 gn2\_06 &  12:36:39.03 &  62:16:41.42 &   41.04 &       - &   123.3 &   119.3 &   119.3 &       - &    0.69 &    0.65 &    0.75 &       - &    22.0 &    21.8 &    21.1 &    28 \\
 gs2\_01 &   3:32:30.94 & -27:43: 4.41 &   74.40 &       - &       - &    71.6 &   116.3 &       - &       - &    0.62 &    0.81 &       - &       - &    21.4 &    20.8 &    26 \\

\hline
$z\sim3.3$ &          &                &        &       &       &       &       &      &      &      &      &      &      &      &      & \\          
 ae3\_04 &  14:19:29.54 &  52:50:22.38 &  293.02 &       - &       - &   119.3 &   119.3 &       - &       - &    0.73 &    0.59 &       - &       - &    22.1 &    21.4 &    31 \\
 co3\_01 &  10: 0:24.87 &   2:12:28.66 &   77.90 &       - &       - &   111.3 &   134.2 &       - &       - &    0.68 &    0.67 &       - &       - &    21.7 &    20.8 &    29 \\
 co3\_04 &  10: 0:24.56 &   2:21:43.56 &   87.95 &       - &       - &   119.3 &   119.3 &       - &       - &    0.73 &    0.71 &       - &       - &    22.2 &    21.4 &    29 \\
 co3\_05 &  10: 0:24.03 &   2:24:56.96 &   78.05 &       - &       - &   115.3 &   119.3 &       - &       - &    0.82 &    0.56 &       - &       - &    21.7 &    21.6 &    30 \\
 gn3\_06 &  12:36:40.56 &  62:16:43.02 &   42.36 &       - &       - &   119.3 &   119.3 &       - &       - &    0.58 &    0.66 &       - &       - &    21.9 &    20.9 &    24 \\

\hline
   Total &     &     &         &       253 &      1626 &      2216 &      1947 &         &         &         &         &         &         &         &         &     591 \\

\hline
\hline
\end{tabular}
\tablecomments{Masks ud1\_01, gs2\_01 and co3\_01 were observed using an ABBA dither pattern in all bands. All other masks were observed using the ABA\arcmin B\arcmin\ dither pattern.}
\label{tab:summary}
\end{table*}

We observe B8-A1\,V stars at least two times throughout the night at
similar airmasses to those of the science observations, in order to
derive a response spectrum for each band and correct for telluric
absorption features. During 2012B and 2013A, telluric stars were
observed with a 3-bar slit in the center of the CSU. However, for the
H and K bands, the telluric spectrum acquired in this manner does not
cover the entire wavelength range targeted for all objects. To address
this problem, we used a new mask configuration in 2014A, which has two
three-bar slits that are offset in X-direction, with an alignment box
in the center\footnote{For readers who are interested in using this
  configuration when obtaining telluric star spectra, the associated
  mask name is ``long2pos''}. Together, these slits fully cover the
targeted wavelength range.  Dome flats are obtained for each mask and
each filter as well. In the K-band we also obtain Neon and Argon arc
observations, which are needed for the wavelength calibration at the
long-wavelength edge of the band.

\subsection{First Data}


The MOSDEF survey is planned to be executed over four spring
semesters, from 2013 until 2016. Including the pilot run on 22-24
December 2012, the MOSDEF survey has been allocated 24 nights in
2012-2014, with 23 additional nights planned in 2015 and 2016. Of the
24 nights in 2012-2014, we obtained usable data during 14.5
nights. Two nights were lost due to technical problems; another 7.5
nights were lost due to poor weather conditions, during which no
useful observations were obtained.

Table~\ref{tab:summary} provides an overview of the masks observed, of
which 6 target the lower redshift regime, 10 the middle redshift
regime, and 5 the higher redshift regime. The mask names include the
target field (ae: AEGIS, co: COSMOS, and gn: GOODS-N), the redshift
interval (1: $1.37 \le z\le 1.70$, 2: $2.09 \le z\le 2.61$, and 3:
$2.95\le z\le 3.80$), and the pointing number (see
Figure~\ref{fig:fields}). For our pilot observing run in December 2012
we also observed two masks in additional CANDELS/3D-HST fields: one in
GOODS-S (gs) and one in UDS (ud). The number of targeted galaxies per
mask ranges from 24 to 33, with an average of 28 galaxies. The 21
masks observed to date have resulted in 591 2D galaxy spectra. This
number does not include the 122 additional objects that
serendipitously fell on slits and for which 1D spectra were extracted
(see Section~\ref{sec:ext}).

Table~\ref{tab:summary} also gives the mask parameters for each mask,
which include the right ascension (RA), declination (DEC), and the
position angle (PA) of the CSU. The slits are tilted by 4$^{\circ}$
relative to the CSU. Due to the fixed angle of the slits, we observe
galaxies at random orientations compared to their major axes. Because
of the random slit orientations and the small sizes of most galaxies,
only $\sim$25\% of the galaxies exhibit resolved velocity information
in their emission lines (S. Price et al, in preparation). The actual
mask parameters differ slightly from the pointings presented in
Figure~\ref{fig:fields}, as we allow the PA to vary by $\pm 5^{\circ}$
and the mask center to vary by $\pm 10\arcsec$ in the $x$ and $y$
directions when finding the optimal mask configuration. However, as
the optimal parameters for one pointing differ per redshift interval,
we only show the nominal parameters for each pointing in
Figure~\ref{fig:fields}.

In addition to the mask parameters, Table~\ref{tab:summary} also gives
the integration time, seeing and depth of all masks and filters (see
Section~\ref{sec:red} for the measurement procedure). The seeing values
are derived from the profiles of the slit stars in the final reduced
and combined frames. However, as in our reduction procedure poor
weather frames have a lower weight, the actual seeing variations are
larger than the spread in Table~\ref{tab:summary} and range from
0\farcs4-1\farcs6. Thus, the values listed for the seeing are the
effective seeing measurements for the corresponding mask. The depth is
also measured in each final reduced frame, as explained in
Section~\ref{sec:cont}.

While our nominal integration times are 1 hour per filter for the
$z\sim1.5$ masks, and 2 hours per filter for the $z\sim2.3$ and
$z\sim3.3$ redshift masks, the actual integration times deviate in
many cases. This difference is due to scheduling constraints (e.g.,
end of night), the removal of problematic frames, or poor weather
conditions. The total integration time is just over 100 hours. 

During mediocre but still observable weather conditions, priority was
given to dedicated low redshift ``bad weather'' masks: ae1\_05,
co1\_03 and co1\_05. By increasing the integration times, we reached
almost comparable depths to those of the typical masks. However, as the
weather is unpredictable, several ``good weather'' masks were observed
during non-optimal weather conditions as well. Consequently, the depth
for similar exposure times is variable (see Table~\ref{tab:summary}).

To reach our goal of $\sim$1500 galaxies, we plan to observe 7, 17 and
8 more $z\sim1.5$, $z\sim2.3$ and $z\sim3.3$ redshift masks,
respectively. These additional observations will bring the total
number of masks to 13, 27, and 13 for the respective redshift
intervals. For the middle redshift regime we aim to complete all 26
pointings shown in Figure~\ref{fig:fields}, in addition to the
pointing in GOODS-S which is not shown in this figure. For the other
redshift intervals we will observe roughly half of the pointings
indicated in Figure~\ref{fig:fields}.

\section{DATA REDUCTION}\label{sec:red}

\subsection{2D Data Reduction}\label{sec:rec}

The MOSFIRE data were reduced using a custom software package written
in IDL, which produces 2D reduced spectra from raw data in a fully
automatic fashion. In summary, this package removes the sky,
identifies and masks cosmic rays and bad pixels, rectifies the frames,
combines all individual exposures, corrects for the telluric response,
and performs an initial flux calibration. The program relies on a
short parameter file that lists the raw data directory, the target
name (from the starlist), the mask name, the MAGMA directory, and the
filter. For flux calibration one can also specify directories pointing
to the response curves and the photometric catalog from which the
targets were selected (see Section~\ref{sec:cal}).

We start by identifying all relevant frames, using the header
information and the parameter file. In addition to all science and
calibration frames, we also identify the sky frames to be used for
each science frame. This identification is based on the dither offset
and the time difference between two consecutive frames. For example,
the first frame of a sequence only has one sky frame, while a central
exposure has two sky frames for an ABA\arcmin B\arcmin\ dither
pattern. For an ABBA dither pattern there is only one sky frame for
each science frame. The headers also indicate which pair of bars belong
to which slit. Finally, using the MAGMA files we assign a 3D-HST ID number
to each slit, identify which of the slits target stars, and
obtain an initial estimate of the wavelength calibration based on the
horizontal position of the slit in the CSU.

Using the information collected in the first step we perform an
initial sky correction on each raw science frame. If only one sky
frame is identified, we simply subtract the sky from the science
frame. In cases for which two sky frames are identified, we subtract
the average of the two adjacent sky frames. We also make a master sky and
master flat frame for each filter and mask. The master sky frame is
constructed by taking the median of all raw science
frames\footnote{There is typically no flexure, and thus sky lines stay
  at the same position in all raw frames for a specific mask and
  filter} and will be used for the wavelength
calibration. The master flat frame is constructed by averaging the
individual dome flat exposures. For each science exposure we 
construct a master mask, in order to mask bad pixels and cosmic rays
while combining the individual frames. The bad pixel map is adopted
from the official MOSFIRE reduction pipeline, constructed by
N. Konidaris. A cosmic ray map is constructed for each science
frame separately, by running the L.A.Cosmic routine
\citep{PvanDokkum2001} on the sky-subtracted science frame. We then
combine the cosmic ray masks with the bad pixel map to make a master
mask for each science frame.

Using the master flat we correct all sky-subtracted science frames for
sensitivity differences. We do not correct the flat for the response
in wavelength direction, as it differs slightly among the different
slits. By retaining the flat response, we can later correct for these
differences (see Section~\ref{sec:cal}). 

Next, we derive all information needed to rectify the raw
spectra. This procedure consists of several steps. However, for an
optimal reduction, it is crucial to resample the data as few times as
possible. Thus, we combine the results of all steps into one
transformation, which rectifies each raw frame to the same reference
coordinate system. The reference coordinate systems has the same
dispersion and pixel scale (0\farcs 1799 per pixel) as the raw frames,
with the wavelength and spatial direction oriented along the $x$ and
$y$ directions, respectively.

In the first step of the rectification procedure we use the master
flat frame to identify the edges of each spectrum. We encounter two
challenges regarding the edge tracing. First, when two neighboring
slits are very close in horizontal position, the dividing line between
the two spectra cannot be accurately determined. In this case the slit
edges are calculated using the top edge of the upper slit and the
bottom edge of the lower slit. Second, the bottom slit on each mask
has been cut off, and thus we cannot accurately measure the bottom
edge. Thus, the rectification for this spectrum may not be optimal.

In the second step of the rectification procedure, we straighten all
spectra in the master sky frame using the slit edges and measure the
positions of bright sky lines. To assign a wavelength to each position
and thus solve for $\lambda(x,y)$, we fit a 2D polynomial to the $x$
and $y$ positions of the sky lines using the IDL function {\tt SFIT}
with a maximum degree of 3 for both dimensions combined\footnote{We
  fit the function $\lambda(x,y) =
  a+by+cy^2+dy^3+ex+fxy+gxy^2+hx^2+ix^2y+jx^3$, with $x$ the
  horizontal and $y$ the vertical position of the sky lines in the
  spectrum rectified using  the slit edges. No term has a combined $x$
  and $y$ power higher than 3.}. This step is automated, as the
horizontal position of a slit provides us with a rough estimate of the
wavelength solution. For the K-band exposures we use Argon and Neon
arc frames as well, as there are no bright sky lines beyond
2.3\,$\mu$m. The master sky and arc frames are slightly shifted with
regard to each other, and while finding the wavelength solution, we
fit and correct for this offset. The combination of the slit edges and
the 2D polynomial fit gives us the wavelength solution for each pixel
in each original frame.

\begin{figure}
  \begin{center}  
  \includegraphics[width=0.42\textwidth]{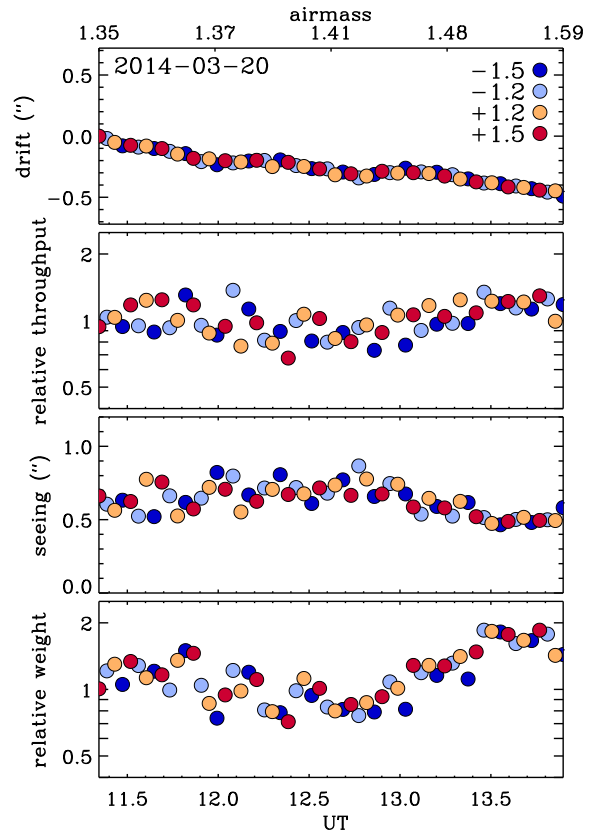}        
 
  \caption{Slit star statistics for the J-band observations of mask
    gn2\_04, as a function of universal time (UT) and airmass. Each
    dot represents an individual frame, and each color indicates a
    different dither position. For each mask and each frame we model
    the slit star profile with a Gaussian in a preliminary rectified
    frame. From this Gaussian we determine the relative position
    (drift), throughput, seeing, and weight. In the top panel we show
    the position compared to the first frame corrected for the dither
    offset. In the 2nd and 3rd panel from the top we show the relative
    throughput (i.e., integral under the Gaussian fit) and the FWHM of
    the seeing, respectively. The bottom panel presents the weight,
    which is the maximum flux of the Gaussian fit, and thus
    proportional to the throughput and the inverse
    seeing. \label{fig:slitstar}}

\end{center}  
\end{figure}

\begin{figure*} 
  \begin{center}             
  \includegraphics[width=0.98\textwidth]{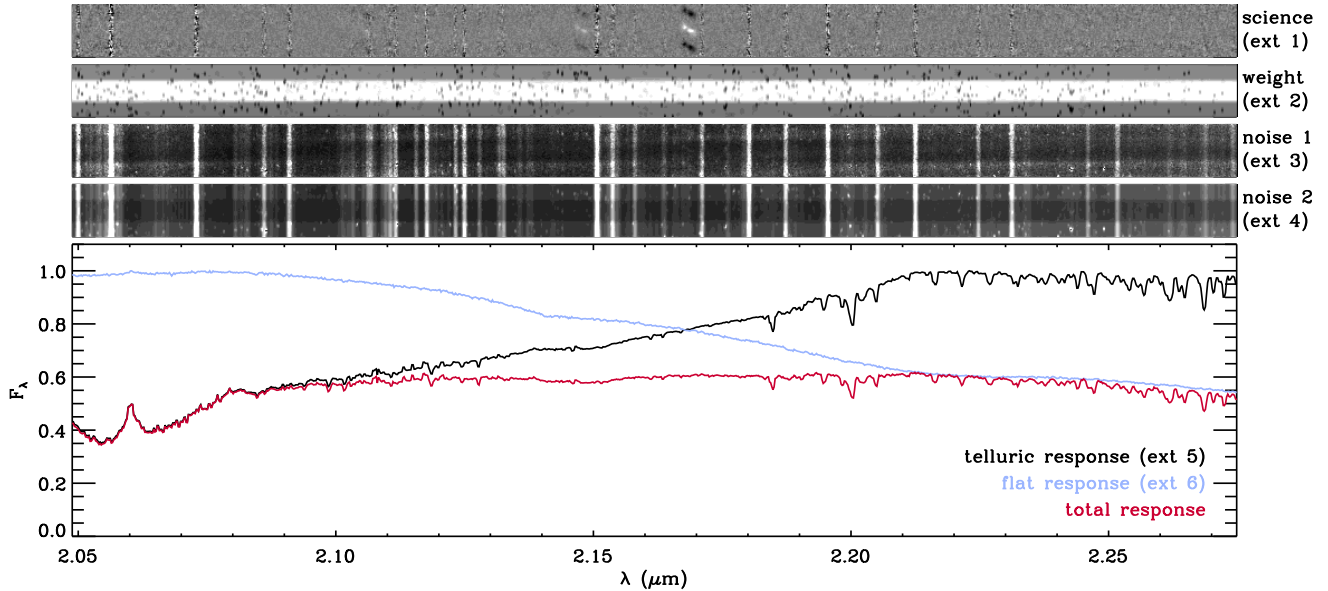} 

 \caption{Overview of the MOSDEF 2D reduction pipeline data products
   for a $z\sim3$ target in the K-band. The top panel shows the 2D
   science frame, in which two \oiii\ lines and a faint \hb\ line are
   visible in white. The negative emission lines in black are the
   result of using offset science frames as sky. The second panel from
   the top shows the weight frame with light pixels having larger
   weight. The gray horizontal bands indicate the regions that have
   been targeted by only half of the exposures. Dark spots indicate
   bad pixels and/or cosmic rays. The third panel from the top shows
   the noise frame constructed from variations between the separate
   science exposures. The fourth panels shows the noise frame
   constructed from the total number of counts, the gain and the
   readout noise. We use this frame to derive our 1D error
   spectra. The bottom panel shows the 1D flat (blue) and telluric
   (black) response spectrum. Combined they form the total response
   spectrum (red). Only a fraction of the total K-band wavelength
   range is shown.\label{fig:spec}}

\end{center}       
                     
\end{figure*}

The last piece of information required to rectify the spectra is the
relative offset of the frames. The dither stored in the header is used
as an initial guess of the position of the target. Using this initial
guess, we find the position of the slit star, which is used as the
real offset. For this measurement, we first rectify the sky-subtracted
spectrum of the brightest slit star to the reference coordinate system
using the wavelength solution derived in the previous step in
combination with the assigned dither positions. Next, we measure the
position of the star in the rectified frames, by fitting the profile
with a Gaussian. In Figure~\ref{fig:slitstar} we show the y-position
(i.e., spatial position) of the slit star for all exposures of one
example mask and filter (gn2\_04, J). It is clear that there is a
systematic and monotonic drift of $\sim$0\farcs2 ($\sim$1 pixel) per
hour. This drift occurs in nearly all sequences and is on average
$\sim$1 pixel per hour. The strength and direction of the drift vary
with field and airmass. We correct all dither positions for this
drift.

For each spectrum in each science exposure we derive the
transformations between the raw and reference frame using the
drift-corrected dither positions and the wavelength solution. By
combining all transformations in one step, we resample our data only
once, thus avoiding further smoothing and noise correlations. Using
these transformations, we rectify all sky-subtracted and
sensitivity-corrected spectra to the reference coordinate system. Hence, this
is the only step during which our data are resampled. We also resample
the combined masks and set all pixels that are affected $>$5\% by a
bad pixel or cosmic ray to 0. We combine the resampled science masks
with the resampled masks for the sky frame(s). We remove any
additional sky from the science spectra by subtracting the median at
each wavelength.

The slit star provides a seeing and throughput measurement for each
science exposure. We use these measurements to determine the relative
weight of the different frames. We take the maximum flux of the
best-fit Gaussian to the slit star profile as the weight factor. By
using the maximum, we optimize according to image quality and
throughput, which both contribute to a higher S/N of the final 1D
extracted spectrum. The throughput, seeing, and weight of each
  individual frame for one example mask are shown in the lower three
  panels of Figure~\ref{fig:slitstar}.

Finally, we combine all rectified, sky-subtracted, and
sensitivity-corrected frames according to their weights, while masking
bad pixels and cosmic rays, as described by 
\begin{equation}
\bar n_{x,y} = \frac{\sum_{i=1}^N w_i m_{x,y,i} n_{x,y,i}}{\sum_{i=1}^N w_i m_{x,y,i}}
\end{equation}
with $n_{x,y,i}$ the number of counts of pixel $(x,y)$ in the
rectified, sky-subtracted, and sensitivity corrected science frame
$i$, $\bar n_{x,y}$ the weighted mean of the count level of all
frames at pixel $(x,y)$, $w_i$ the weight of science frame $i$,
$m_{x,y,i}$ the mask value (1: included; 0: excluded) at pixel $(x,y)$
of science frame $i$, and $N$ the number of science frames.

In order to construct a noise frame for each reduced science spectrum,
we rectify the original raw science and corresponding (non-flat
fielded) sky frame as well. We make a noise frame for each individual
science exposure by combining the counts of the rectified science and
sky frames and adding the readout noise in quadrature. If there is one
sky frame, we simply add the counts of the science and the sky frame,
and multiply the readout noise of an individual frame by
$\sqrt{2}$. We use the following expression to derive
$\sigma_{x,y,i}$, the noise at pixel $(x,y)$ of individual science
exposure $i$:
\begin{equation}
\label{eq:noise_onesky}
\sigma_{x,y,i} = \frac{\sqrt{Gs_{x,y,i} + Gs_{x,y,i\pm1} + 2R^2}}{G}
\end{equation}
with $s_{x,y,i}$ the total number of counts of pixel $(x,y)$ in the
rectified, non-sky-subtracted science frame $i$, $s_{i\pm1,x,y}$ the
number of counts in pixel $(x,y)$ in either the previous
($s_{i-1,x,y}$) or next ($s_{i+1,x,y}$) science frame used as sky, $G$
the gain, and $R$ the readout noise.

In cases where the sky frame is constructed of two surrounding frames
in a ABA\arcmin B\arcmin\ dither sequence, the sky noise will go down
by $\sqrt{2}$. As we take the average of two sky frames, the sky noise
(in electrons) becomes $\sqrt{G (s_{x,y,i-1}+s_{x,y,i+1}) / 4}$. For
the readout noise we add in quadrature the readout noise of the
science frame ($R$) and the combined readout noise of the sky frame
($R/\sqrt{2}$), which becomes $\sqrt{3/2}R$. By adding the background
noise of the science frame, the background noise of the combined sky
frame, and the total readout noise, we get the following expression:
\begin{equation}
\label{eq:noise_twosky}
\sigma_{x,y,i} = \frac{\sqrt{Gs_{x,y,i} + G (s_{x,y,i-1}+s_{x,y,i+1}) / 4 + 3R^2 / 2}}{G}
\end{equation}

We construct the final noise frame by adding the individual noise
frames in quadrature, while taking into account their weights. For
each individual pixel $(x,y)$ we use the following expression:
\begin{equation}
\sigma_{x,y} = \frac{\sqrt{\sum_{i=1}^N \left( w_i m_{x,y,i} \sigma_{x,y,i} \right)^2}}{\sum_{i=1}^N w_i m_{x,y,i}}
\end{equation}
Finally, we correct the noise frame for the flat
response, using the rectified flat frame.

We make a second noise frame, based on the variations between the
rectified science exposures. This noise frame is based on resampled
data and thus we correct for random 2D resampling by multiplying the
noise frame by a factor of 1.52\footnote{This factor is derived by
  comparing the original noise of a frame to the noise after we
  randomly resample in both directions, while keeping the pixels the
  same size. We repeat this procedure 10,000 times, and derive the
  average factor by which the noise has decreased}. We use the
following expression:
\begin{equation}
\sigma_{x,y} = 1.52 \sigma_{x,y,s}\frac{ \sqrt{\sum_{i=1}^N \left(w_i m_{x,y,i}\right)^2}}{\sum_{i=1}^{N} w_i m_{x,y,i}}
\end{equation}
with $\sigma_{x,y,s}$ the sample standard deviation:
\begin{equation}
\sigma_{x,y,s} = \sqrt{\frac{\sum_{i=1}^N m_{x,y,i} \left(n_{x,y,i} - \bar n_{x,y}\right)^2}{N_{x,y}'-1}}
\end{equation}
with $N_{x,y}'$ the number of frames with non-zero weight at pixel $(x,y)$.

Finally, we construct a weight map, by combining the weights and
rectified masks for each science frame.
\begin{equation}
W_{x,y} = \frac{\sum_{i=1}^N w_i m_{x,y,i}}{\sum_{i=1}^N w_i}
\end{equation}
with $W_{x,y}$ the total weight at pixel $(x,y)$. 

The different reduction products are presented in
Figure~\ref{fig:spec}. We use the noise spectrum derived using
Equations (2)-(4) for our 1D error spectrum (extension 4 in
Figure~\ref{fig:spec}). Nonetheless, in Section~\ref{sec:noise} we
show that the different error spectra are consistent with each other.

\subsection{Calibration}\label{sec:cal}

All spectra are calibrated for the relative response using telluric
standards. During most nights we observed telluric standards at
similar airmasses as the science observations. These spectra are
reduced using the same method as that applied to the science
masks. Similar to the science observations, we do not remove the flat
response in the reduction. Thus, the overall flat response is canceled
once the science spectra are divided by the telluric response
spectrum. However, by keeping the response in the flat, we correct for
the small difference in sensitivities as a function of wavelength
among the different slits.

We compare the observed telluric stellar spectra with the intrinsic
spectra for the corresponding spectral type. The spectral types of the
telluric standards range from B8\,V to A1\,V. These stars have Balmer
and Helium absorption lines, which may differ in line width or depth
from the telluric standards we observed. Thus, in both the observed
and the intrinsic spectra we interpolate over stellar absorption
features. We derive a response spectrum by dividing the observed by
the intrinsic stellar spectrum of the same spectral type. For the
interpolated regions, we multiply the response spectrum by a
theoretical sky absorption spectrum for the corresponding airmass.

\begin{figure} 
  \begin{center}           
     \includegraphics[width=0.46\textwidth]{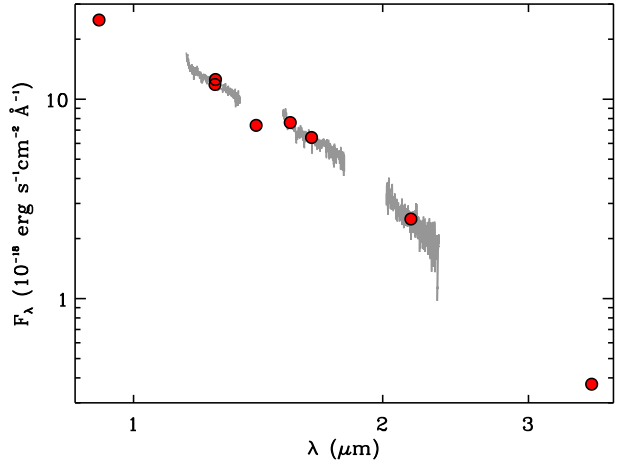} 

     \caption{Comparison of the MOSDEF J, H, and K spectra of a slit
       star with the photometry from 3D-HST ($H_{F160W}=19.45$). The
       slit star spectra are scaled to match the 3D-HST
       photometry. All galaxies within the same mask are calibrated
       using this same normalization factor. A second correction is
       applied to take into account that slit losses depend on the
       spatial extent of the galaxies. This figure further illustrates
       the excellent agreement between the photometric and
       spectroscopic shapes.  \label{fig:scaling}}   

  \end{center}                            
\end{figure}

To construct the response spectrum for a specific mask and filter, we
combine multiple tellurics at similar airmass observed over different
nights, to match the effective airmass of the final science
frame. There are several reasons why we adopt this approach, instead
of using only one response spectrum for a specific night. First,
telluric standards are not available for all nights, and, if they are
available, the airmass match is often not optimal. Second, some
telluric spectra are noisy, which would increase the noise of our
galaxy spectra. Third, the telluric spectra taken in 2012 and 2013
do not have full coverage in the H and K bands (see
Section~\ref{sec:obs}). Finally, we do not find substantial difference
in the atmospheric absorption features at a given airmass between
different nights. When combining the response spectra of different
nights, we take into account differences in the flat response, as a
flat lamp change took place on 13 February 2014.

For the absolute flux calibration we make use of the slit star. First,
we extract a 1D spectrum using the same optimal extraction method that
we use for the galaxy spectra (see Section~\ref{sec:ext}). We derive a
scaling factor by comparing this slit star spectrum with the
photometry in the 3D-HST photometric catalogs. As most slit star
spectra are only partially covered by any photometric filter, we do
not integrate the spectra with the filter response curve of the
closest 3D-HST photometric band. Instead, we fit the 3D-HST photometry
and scale the spectrum to the fit. We only use the corresponding and
surrounding filters to ensure that the fit perfectly matches the
photometry, and we assume simple blackbody shape. We calibrate all
spectra within a mask using this normalization factor. The
normalization factors, which include the slit-loss correction for a
point source, range from $0.8-5.9\times 10^{-17} \rm ~erg~ cm^{-2}~
\AA^{-1}~ cts^{-1}$. In Figure~\ref{fig:scaling} we show the scaled 1D
spectrum of the slit star of mask gn2\_06 in combination with the
3D-HST photometry used for the scaling.

\subsection{Slit Loss Corrections}\label{sec:slitloss}

A crucial step in using the MOSDEF spectroscopy to calculate line flux
ratios and absolute luminosities is to account for the loss of flux
outside the slit apertures. As explained in the previous section, all
spectra on a given mask for a given filter are scaled by a
normalization factor computed by comparing the slit star spectrum with
the 3D-HST photometry. This procedure accounts for both the conversion
of counts to flux, as well as the slit loss assuming that the galaxies
are unresolved point sources. However, most of the galaxies are in
fact resolved based on an estimate of their sizes from the {\em HST}
F160W images and the typical seeing of our
observations.

Thus, we adopt the following procedure to better account for the loss
of flux outside the slit aperture for each galaxy. First, we extract
an F160W postage stamp of the galaxy from the CANDELS F160W imaging
\citep{RSkelton2014}, and use the SExtractor \citep{EBertin1996}
detection segmentation map\footnote{The segmentation map indicates
  which pixels contain flux from objects and has non-zero values where
  objects are detected. Pixels without detections have a zero value.} 
to mask out all pixels in the postage stamp belonging to other nearby
sources, retaining pixels that sample the background. The masked out
pixels are replaced by noise that is calculated from the average and
standard deviation of the background pixel values. Second, we smooth
the postage stamp with a Gaussian kernel with FHWM = $\sqrt{{\rm
    FWHM_{\rm seeing}}^2 - {\rm FWHM_{\rm F160W}^2}}$, where
FWHM$_{\rm seeing}$ is the seeing derived from the Gaussian fit to the
profile of the slit star observed on the same mask and in the same
filter (see Section~\ref{sec:rec}) and FWHM$_{\rm F160W}$ is the FWHM
of the F160W PSF. Third, we fit the smoothed image of the galaxy with
a two-dimensional elliptical Gaussian which is allowed to rotate
freely to obtain the best fit. Fourth, the modeled profile is rotated
into the frame of reference of the slit, taking into account the
position angle of the slit. Fifth the elliptical Gaussian is
integrated within the slit boundaries to arrive at the fraction of
light contained within the slit. Finally, each spectrum is multiplied
by the ratio of the included fraction of light measured for the slit
star (assuming a circular Gaussian) and that measured for the
elliptical Gaussian. This last step corrects each galaxy spectrum for
the additional light lost outside of the slit, relative to the slit
star.

\begin{figure} 
  \begin{center}           
     \includegraphics[width=0.48\textwidth]{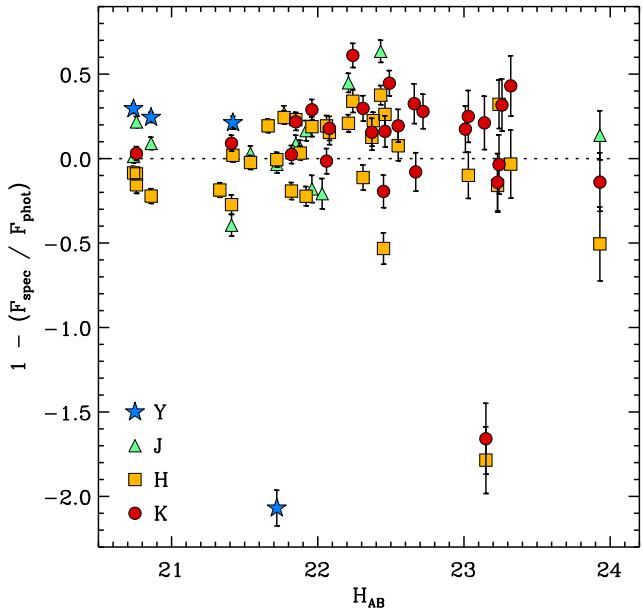} 

     \caption{Comparison of the photometric to spectroscopic flux as a
       function of $H_{\rm F160W}$ magnitude for galaxies with
       continuum detections. All MOSDEF spectra are calibrated using a
       scaling factor derived by comparing the spectrum and photometry
       of the slit star, in combination with a slit loss
       correction. Hence, the photometric fluxes of the galaxies are
       not used to calibrate the spectra, and thus a comparison with
       the spectroscopic fluxes gives an estimate of the uncertainty
       on the flux calibration. This comparison indicates that the
       uncertainties in flux calibration are $\approx 16\%$, with a
       bias of less than $18\%$. \label{fig:slitloss}}   

  \end{center}                            
\end{figure}

The effectiveness of this procedure is tested by first calculating the
median fluxes from the best-fit SEDs of the continuum-detected
galaxies  (where the median is computed between wavelengths covered by
our actual  spectra), and then comparing these median SED-inferred
fluxes with the median spectroscopic fluxes (see
Figure~\ref{fig:slitloss}). Taking into account the measurement
errors, this comparison indicates that the uncertainties in flux
calibration are $\approx 16\%$, with a bias of less than
$18\%$. Similarly, for the line ratios spanning different filters
(e.g., \ha/\hb, \oiii/\oii), we find a random uncertainty of $18\%$
and a bias of less than $13\%$. 

We caution that the slit loss corrections are based on rest-frame
optical continuum emission, and thus the total emission line fluxes as
derived in Section 4 may be over or underestimated, depending on
whether the line emission is more or less concentrated than the
continuum emission. One possible method to address this issue is to
correct the line emission separately for slit losses using HST imaging
in the rest-frame UV, which is more sensitive to star-forming
regions. However, the use of bluer bands suffers from other
complications, such as patchy dust attenuation, and thus it will not
necessarily result in more robust slit loss corrections.

\subsection{Noise Properties}\label{sec:noise}

To quantify the significance of our results, it is crucial that we
understand the noise properties of our observations. As described in
Section~\ref{sec:rec}, we construct two independent noise frames for
each reduced mask. The first noise frame is derived from the number of
counts, gain, and readout noise level of the detector (Equations 2-4,
extension 4), and the second one is based on the variations between
the rectified individual frames (Equations 5-6, extension 3). As both
noise frames are derived using indirect methods, we use a third
independent method, which is based solely on the reduced spectra,
without using intermediate data products. For each of these three
methods we make a 1D noise spectrum. For a fair comparison we only
consider the wavelength range that is covered by all slits and masks
for a specific filter (Y: 9700-10900\,\AA, J: 11700-13100\,\AA, H:
15400-17200\,\AA, K: 20400-22900\,\AA).

The third method is based on the variations of extracted 1D spectra in
empty, yet fully exposed regions of the reduced 2D spectra. First, we
take all reduced 2D spectra of a mask  we exclude all regions that are
not fully exposed, by removing the rows that have a normalized median
weight of less than 0.92 (i.e., non-fully exposed regions). Thus, the
gray and black rows in the weight panel of Figure~\ref{fig:spec} are
not included. Second, we bin the 2D reduced spectra in the wavelength
direction by 5 pixels, to remove the effect of resampling in the 2D
reduction. Third, we extract spectra in the empty regions (i.e.,
avoiding objects) using the same optimal extraction method as for our
real spectra and a profile with a FWHM of 0\farcs 6, to represent
typical seeing values for our observations. We extract as many
independent empty spectra as possible across all slits on the
mask. Each mask allows roughly 80 such empty apertures. Finally, we
use the standard deviation at each wavelength as an estimate of the 1D
noise spectrum.

\begin{figure} 
  \begin{center}             
  \includegraphics[width=0.48\textwidth]{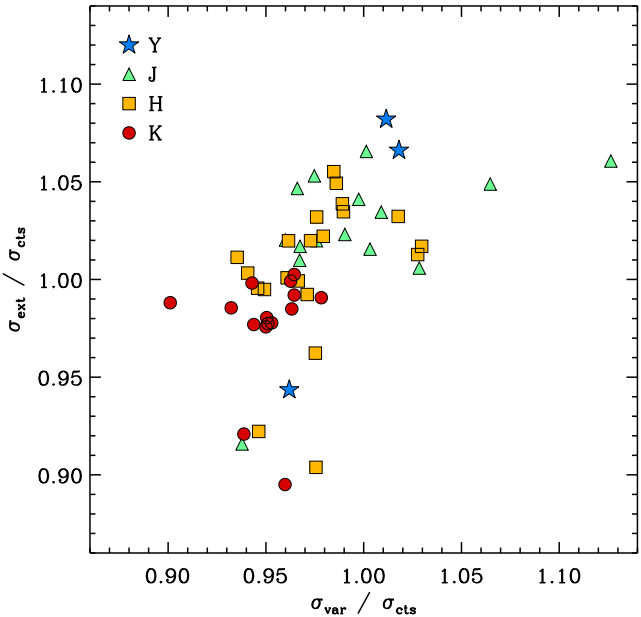} 

 \caption{Comparison of three different noise measurements, as derived
   through independent methods. For all noise measurements, we
   determine the median noise level over the wavelength range covered
   by all spectra for a specific filter (Y: 9700-10900\,\AA, J:
   11700-13100\,\AA, H: 15400-17200\,\AA, K: 20400-22900\,\AA). The
   $x$-axis shows the ratio of the noise as derived from the
   frame-to-frame variations ($\sigma_{\rm var}$) to the noise based
   on total number of counts combined with the gain and readout noise
   ($\sigma_{\rm cts}$). On the $y$-axis we show the ratio of the noise
   as derived by extracting spectra in empty regions ($\sigma_{\rm
     ext}$) to $\sigma_{\rm cts}$. The mean of $\sigma_{\rm
     ext}/\sigma_{\rm cts}$ for all masks and filters is 1.00, while
   the mean of $\sigma_{\rm var}/\sigma_{\rm cts}$ is 0.97. \label{fig:noise}}

\end{center}       
                     
\end{figure}

To compare the results of this third method with the two different 2D
noise spectra produced by the reduction procedure, we use the same
``empty and fully exposed'' regions. But first, similar to the reduced
2D science spectra, we bin the 2D noise spectra in wavelength
direction by 5 pixels as well, by taking the square root of the summed
variance. Next, we (optimally) extract a 1D noise spectrum from the
binned noise frame for each empty region (corresponding to the same
regions used to extract the empty spectra in the science spectra),
assuming the same profile. Finally, we take the average of the 1D
noise spectra for all empty regions to construct the mean 1D noise
spectrum. Using this method, we make a 1D noise spectrum for both the
noise frames in extension 3 and extension 4. To compare the different
1D noise spectra, we assess their ratio as a function of
wavelength. As the ratios show no trend with wavelength, we derive a
median value for the ratios for each mask over the wavelength regions
covered by all spectra (defined in the first paragraph of this
section). The results of this test are presented in
Figure~\ref{fig:noise}.

Figure~\ref{fig:noise} illustrates that the three different noise
measurements agree very well. The median ratio between the empty
aperture and sky noise (extension 4, Equations 2-4) method for all
different masks and filters is 1.0. The median ratio between the frame
variation (extension 3, Equations 5-6) and sky noise method is
0.97. These ratios are robust against different aperture sizes. There
is a slight difference between the different filters, with the K-band
yielding a relatively lower noise level for the empty aperture method
compared to the other methods. The exact reason for this difference is
not well understood. Hereafter, we use the sky and readout noise
(extension 4) as the noise spectra in our analysis.

\subsection{Assessment}

Our observational and reduction procedures contain several steps that
differ from the conventional procedure to obtain and reduce near-IR
spectroscopic data. Here, we assess the improvement by these steps, by
turning off the corresponding features in the reduction procedure. We
show that the improvement due to an individual step is generally only
a few percent. However, the inclusion of all steps together may lead
to an improvement of up to 25\% in the total S/N of the 1D reduced
spectra. 

The first step we assess is the dither sequence. We use an ABA\arcmin
B\arcmin\ dither sequence, with a distance between A and B of 2\farcs
7, and a distance between A and A\arcmin\ (and B and B\arcmin) of
0\farcs 3. This observing sequence has several advantages over the
commonly used ABBA or ABAB dither pattern. The first advantage is that
different parts of the detector are used, and thus bad pixels are
spread out over multiple exposures. The second advantage is that we
can use two surrounding frames as sky, which lowers the noise in the
sky frame by a factor of $\sqrt{1/2}$. In Appendix A we show that the
reduction of noise in the sky frame by a factor of $\sqrt{1/2}$ leads
to a reduction of the noise in the sky-subtracted science frame by a
factor of $\sqrt{3/4}$, compared to using only one sky frame. The ABAB
dither sequence also allows the use of two sky frames and thus a lower
noise level in an individual sky-subtracted frame. However, when
adding the sky-subtracted frames this dither sequence is effectively
similar to an ABBA dither sequence with one sky frame (see Appendix
A). For the final 1D spectrum the improvement in S/N is optimal if the
distance between A and A\arcmin\ is larger than the extraction
aperture. 

\begin{figure} 
  \begin{center}           
     \includegraphics[width=0.48\textwidth]{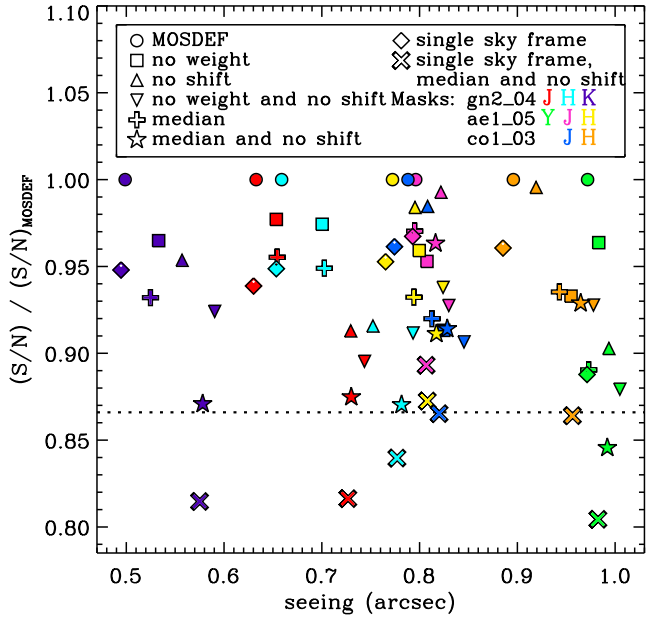} 

     \caption{S/N for alternate reductions, compared to the S/N of the
       official MOSDEF reduction (circle) versus the seeing. We
       explore the improvement due to weighting individual frames
       (square), applying a drift correction (triangle) and correcting
       for both effects (up-side-down triangle). We also include a
       median combine (plus) and a median combine without correcting
       for the drift (star). Finally, we assess the improvement due to the
       inclusion of two instead of one sky frame (diamond). The dotted
       line indicates the theoretical difference ($\sqrt{3/4}$) between using
       one or two sky frames. Compared to a median combine with no
       drift correction and one sky frame (cross), our reduction
       increases the S/N by up to $\sim$25\%
       ($(1.00-0.80)/0.80$). \label{fig:reduction_tests}}
  
  \end{center}                            
\end{figure}

To assess the S/N improvement due to the dither sequence in
combination with the multiple sky frames, we reduced several masks,
while only using one of the surrounding frames as sky (A-B, B-A,
A\arcmin-B\arcmin, B\arcmin-A\arcmin, etc.). When considering the S/N
improvement, we use the noise spectra based on the empty aperture
extractions, as this is the most direct and empirical method to
determine the noise. The noise frame based on the count level
(extension 4)  will by definition give an improvement of
$\sim\sqrt{3/4}$ (the difference between
Equations~\ref{eq:noise_onesky} and \ref{eq:noise_twosky}), and thus
cannot be used as an independent test. The noise spectrum based on the
frame-to-frame variations (extension 3) assumes that the extraction
aperture is smaller than the offset between A and A\arcmin, and thus
can neither be used. In Figure~\ref{fig:reduction_tests} we compare
the ratio in S/N between the reduction with a single sky frame
(diamonds) and the MOSDEF reduction as a function of seeing. As
expected based on the small offset between A and A\arcmin\, the
improvement is less than the theoretical value.

Another improvement in our reduction scheme, facilitated by the
multi-object mode, is the simultaneous monitoring of a relatively
bright star in one of the slits. Using this slit star we weight
individual frames, trace the drift, and calibrate our spectra. In
Figure~\ref{fig:reduction_tests} we show the S/N when reducing the
data without weighting the individual frames (squares) or correcting
for the drift (triangles), compared to our standard MOSDEF reduction
pipeline. We also test the use of a median instead of a mean
(pluses). We use the same mask and noise frame as for the dither test,
so we can assess the total effect of all improvements.

For all cases we find that the MOSDEF reduction results in the highest
S/N. For most masks, the lowest S/N is obtained for a median combine
without a drift correction (stars). Compared to this reduction, the
S/N of the MOSDEF reduction is up to $\sim$18\% higher
($(1.00-0.85)/0.85=0.18$). The combined effect of all different
procedures increases the S/N by up to 25\% ($(1.00-0.80)/0.80=0.25$),
compared to a single sky frame, no shift, and a median combine
reduction (crosses). Increasing the exposure time by 50\% would result
in the same S/N increase.

\begin{figure*} 
  \begin{center}           

     \includegraphics[width=0.32\textwidth]{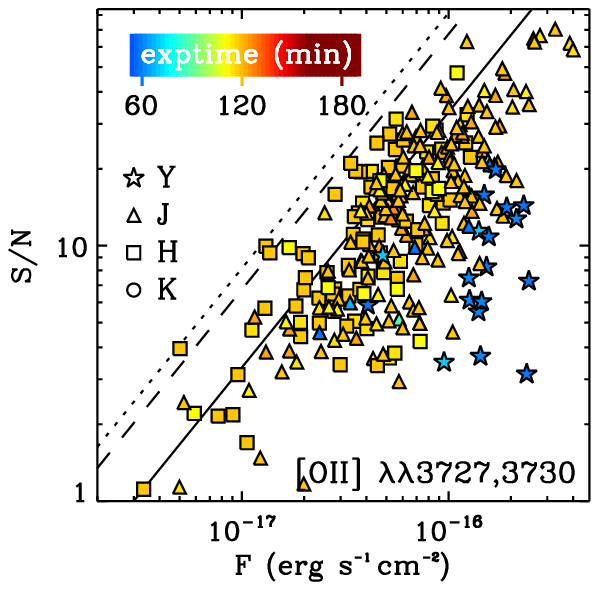}
     \includegraphics[width=0.32\textwidth]{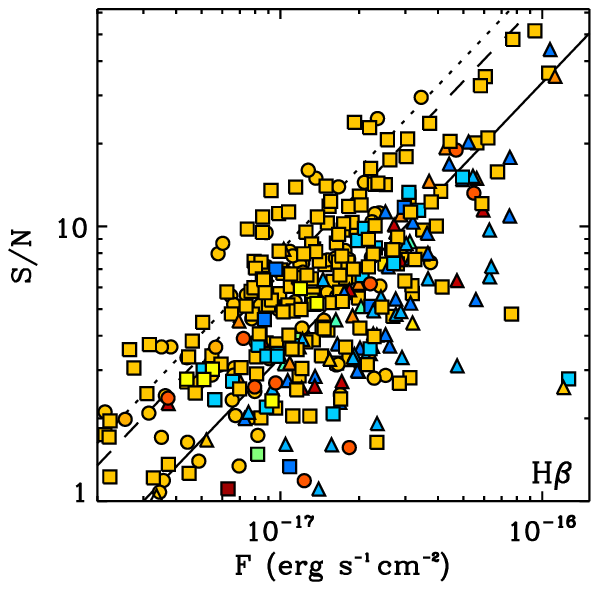}
     \includegraphics[width=0.32\textwidth]{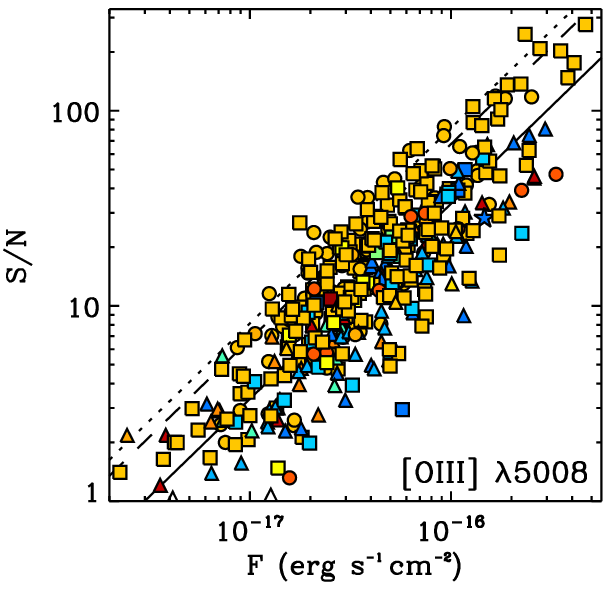}\smallskip\\
     \includegraphics[width=0.32\textwidth]{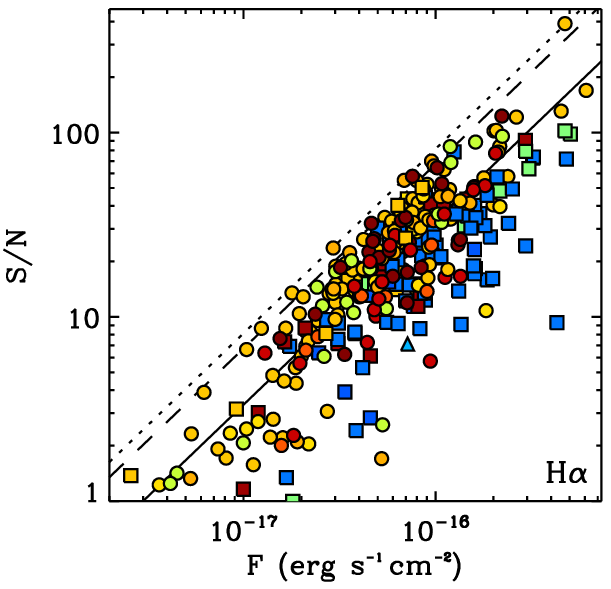}
     \includegraphics[width=0.32\textwidth]{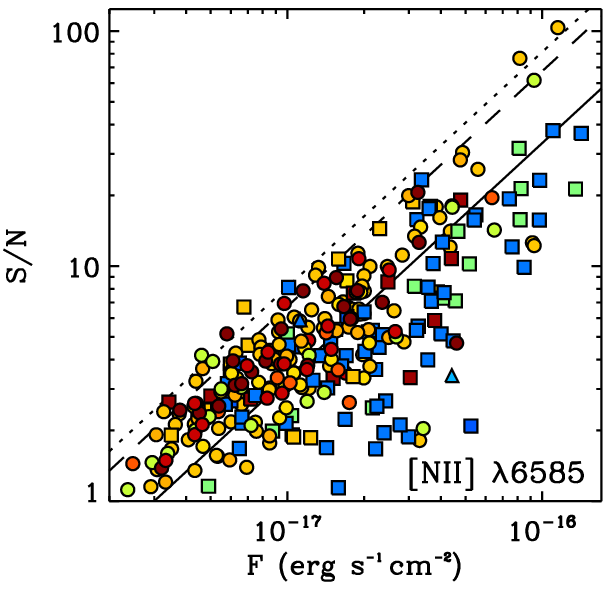}
     \includegraphics[width=0.32\textwidth]{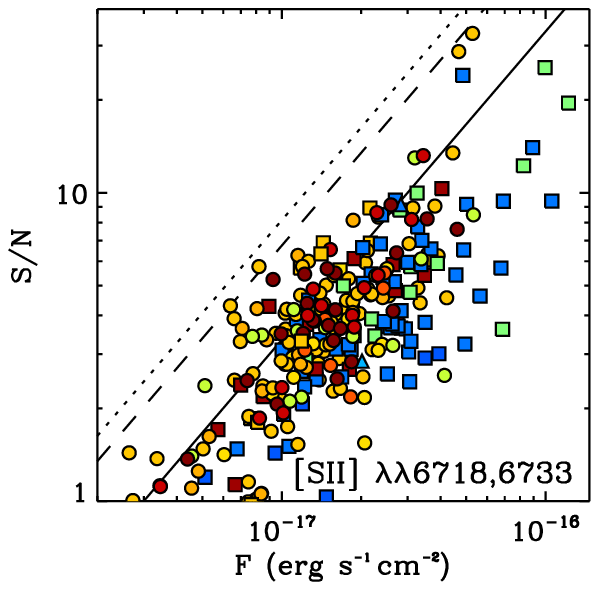}

     \caption{The S/N of emission lines \oii, \hb, \oiii, \ha, \nii,
       and \sii\ vs. the total flux (corrected for slit losses) of the
       line. \ha\ and \hb\ are uncorrected for the underlying
       absorption in this figure. The features are measured in the
       optimally extracted 1D spectra, and serendipitous detections
       are excluded. The color coding reflects the integration time,
       and the symbol indicates the filter in which the line is
       detected. The black solid line shows our typical emission-line
       depth for a 2 hour exposure. The black dotted and dashed lines
       represent our most optimistic depths (i.e., avoiding sky lines
       and bad weather conditions) in 2 hours for H and K,
       respectively. These depths are calculated using the
       \ha\ emission lines. \label{fig:line} }

  \end{center}                            
\end{figure*}

\subsection{Extraction of One-Dimensional Spectra}\label{sec:ext}

1D spectra are extracted by hand using custom IDL software (see Freeman et
al. in preparation for a full
description\footnote{https://github.com/billfreeman44/bmep}). The
extraction program works with output from the custom MOSDEF 2D
reduction pipeline discussed in the previous sections. Both optimally
weighted and unweighted spectra are extracted for each object. The
optimal extraction algorithm is based on \cite{KHorne1986} and is
extended to be able to extract fractions of pixels.

First, the extraction program uses the expected y-position of an
object, as given by the MAGMA output files, to draw a line that
clearly marks the position of the primary object in the 2D
spectra. However, due to the drift correction the expected position is
slightly shifted compared the real position.\footnote{In rare cases,
  an offset was skipped during the observations, resulting in an even
  larger difference between the expected and measured position of the
  spectrum.} We use the position of the slit star in the same mask and
filter to correct the expected position. Using the line at the
expected position we unambiguously identify the primary object. 

Next, the spatial profile of the object is determined by summing only
those columns of the 2D spectra with high S/N in either the continuum
or emission lines. This method provides clean weighting profiles for
the optimal extraction as columns with little or no signal are
excluded. We fit a Gaussian function to the profile to determine the
weighting profile, center, and width of each object. For the
extraction aperture we take twice the FWHM of the Gaussian
function. Finally, 1D spectra are extracted with and without optimal
weighting. We apply this procedure to each object and each filter,
separately. Serendipitous objects are also extracted using the same
method and 89\% of them are identified as objects in the 3D-HST
catalog v4.0 \citep{RSkelton2014}.

In cases where an object has no obvious emission lines or continuum in
the 2D spectrum, a ``blind'' extraction is performed. For objects
with no signal in any band, the blind extraction uses the expected
position of the object, as derived from the MAGMA output file and the
slit star position and uses the same extraction width as the width of
the slit star in each filter. For objects with signal in one or more
bands, the blind extraction uses the average extraction widths and
centers from filters in which signal was detected, corrected for
seeing and offset differences as derived from the slit star profiles
and positions.

\section{ANALYSIS}\label{sec:ana}

In this section we describe the procedure to measure emission line
fluxes and redshifts, present the sensitivities of the line and
continuum emission, and the procedure to derive stellar population
properties from mulit-wavelength photometry in combination with the MOSFIRE
redshifts. We also assess for which galaxies we successfully obtain a
spectroscopic redshift, and how the targeted and spectroscopically
confirmed galaxy samples compare to an H-band limited galaxy sample at
the same redshift.

\subsection{Emission Line Measurements and Sensitivities}

We use a Monte Carlo method to measure emission line fluxes and errors
by perturbing the spectrum of each object by its error spectrum. For
each object, we measure an initial redshift and line-width by fitting
a Gaussian to the highest S/N emission line, which in most cases is
either the \ha\ or \oiii$\lambda 5008$ line. The initial redshift and
FWHM are then used to fit all the other lines of interest. In the
fitting, we allow the observed wavelengths of the lines (as
predicted from the initial redshift) to vary within $\pm 2 \times
(1+z)$\,\AA, and we allow the FWHM to vary within $\approx \pm
0.5$\,\AA\, in the observed frame, excluding values that are lower
than the instrumental resolution. The \oii\ doublet is fit with a
double Gaussian function, and the \ha\ and \nii\ doublet is fit with
three Gaussians. In addition, for all lines we allow for a linear
continuum fit under the Gaussian.  We repeat this procedure using
1000 different realizations of the spectra and calculate the average
line fluxes and dispersions from these realizations. For each line, we
obtain an estimate of the slope and intercept of the continuum under
the line, the observed wavelength, the line flux, and the FWHM.

To more reliably measure fluxes for lines that deviate significantly
from a Gaussian shape, we also simply integrate the flux under each
line using fixed wavelength windows ($f_{\rm window}$).  This second
flux measurement is adopted if: (a) the Gaussian model fit to the
line  deviates from the data by more than 2\,$\sigma$ (as determined
from the error spectrum) in any given pixel within $\pm
3\,\sigma_{\rm w}$  (i.e., line width) of the line center; {\em and}
(b) the absolute value of the difference between $f_{\rm window}$ and
the fitted flux is more than three times the fitted flux error: i.e.,
$|f_{\rm window}-f_{\rm fitted}|$ $>3\,\sigma_{\rm fitted}$.  The
fluxes of the Balmer \ha\ and \hb\ emission lines are corrected for
underlying Balmer absorption using the best-fit stellar population
model, as derived in Section~\ref{sec:starpop}.

In Figure~\ref{fig:line} we show the S/N of the emission lines \oii,
\hb, \oiii, \ha, \nii, and \sii\ vs. the flux of the lines. As the
spectra are corrected for slit losses, the line flux shown here
represents the total line flux of the galaxies, assuming that the line
flux originates from the same region as the stellar continuum as
traced by the F160W images. We only include galaxies for which the
redshift is directly measured from the MOSDEF spectra from either
\ha\ or \oiii, and we exclude serendipitous detections. 

The \oiii\ and \ha\ lines are the strongest emission lines in our
spectra. However, even fainter lines are significantly detected in
most spectra. Within 2 hrs we obtain a 5$\sigma$ line detection for a
typical flux of $\sim1.5\times10^{-17}\rm \,erg\,
s^{-1}\,cm^{-2}$. The line-flux sensitivity is slightly deeper in
H-band, as expected from the MOSFIRE specifications and sky background
level, though the sensitivity difference between filters for our full
set of emission-line measurements is masked by the variation in
observing conditions. In the most optimistic case, when avoiding sky
lines and bad weather conditions, our 5$\sigma$ depth within 2 hours
is $\sim 6.1 \times 10^{-18}\rm \,erg\, s^{-1}\,cm^{-2}$ and
$7.4\times 10^{-18} \rm \,erg\, s^{-1}\,cm^{-2}$ for H and K,
respectively.

For a 5$\sigma$ line detection within 1 hour \cite{CSteidel2014} find
a typical line flux of $3.5\times 10^{-18}$ and $4.5-14 \times
10^{-18}\rm \,erg\, s^{-1}\,cm^{-2}$ in H and K, respectively. In
contrast to \cite{CSteidel2014} our fluxes are all corrected for
slit losses. The mean slit-loss corrections are a factor of 1.69, 1.66,
1.62, and 1.54 for Y, J, H, and K, respectively. Taking into account
the integration time difference and the slit-loss correction, we find a
difference of a factor of 1.5 between our optimistic 5$\sigma$ depth
and the 5$\sigma$ depth found by \cite{CSteidel2014} for the
H-band. In addition, differences in the methods to construct the noise
frames, extract the spectra, measure the emission line fluxes, as well
as in the targeted galaxies (e.g., line widths) and weather conditions
may further contribute to differences in depth.

\begin{figure} 
  \begin{center}           
     \includegraphics[width=0.48\textwidth]{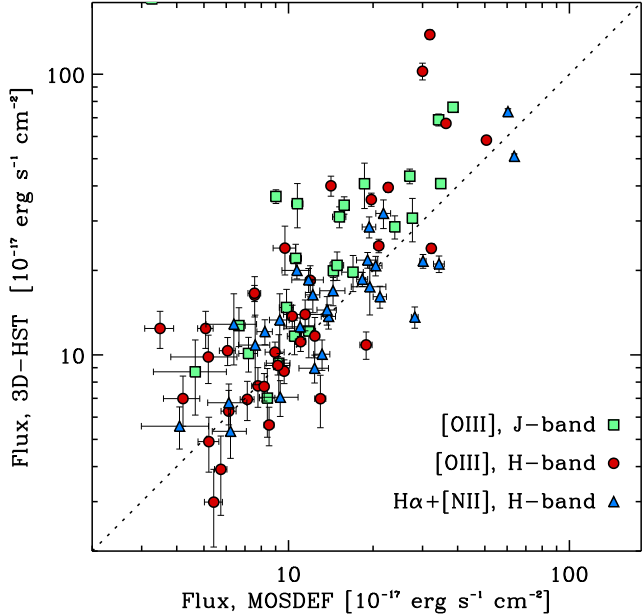} 

     \caption{Comparison of emission line fluxes as measured from
       MOSDEF and 3D-HST. We consider lines with a $\rm S/N>3$ in both
       the MOSDEF and 3D-HST spectra. Due to the lower spectral
       resolution of the 3D-HST grism spectra, we combine \ha\ and the
       two \nii\ lines, and the two \oiii\ lines for MOSDEF. The flux
       measurements for both surveys are corrected to the total flux,
       by scaling the spectra to the 3D-HST
       photometry. \label{fig:line_comp}} 

  \end{center}                            
\end{figure}

We compare the emission line measurements from MOSDEF with those from
3D-HST in Figure~\ref{fig:line_comp}. As the resolution of 3D-HST is
not high enough to deblend close emission lines, we combine the flux
measurements of the unblended lines in MOSDEF as well, for a fair
comparison. In Figure~\ref{fig:line_comp} we only show primary targets
and lines that are detected at 3$\sigma$ in both data sets. \oiii\ is
the sum of the 4960\,\AA\ and 5008~\AA\ lines, and for \ha+\nii, we
add the flux of the \ha\ line (corrected for Balmer absorption)
and the two \nii\ lines at 6550\,\AA\ and 6585\,\AA.

The two datasets agree reasonably well with a median offset and scatter
($\sigma_{\rm NMAD}$) in $1-F_{\rm MOSDEF}/F_{\rm 3D-HST}$ of $\sim$13\%
and 35\%, respectively. The random difference is larger than expected
based on the uncertainties on the individual MOSDEF and 3D-HST line
measurements. This is not surprising, as both uncertainties do not
take into account errors introduced by the absolute flux
calibration. In Section~\ref{sec:slitloss} we find a random
uncertainty of $18\%$ and a bias of less than $13\%$ for the MOSDEF
absolute flux calibration. Furthermore, these values do not include
additional errors on the slit-loss corrections due to the fact that
the line emission may not follow the continuum emission. The 3D-HST
spectra are also scaled using the photometry. This procedure may
introduce additional (systematic) uncertainties on the 3D-HST emission
line fluxes as well.

For galaxies for which the weighting profile is determined from line
emission {\it and} for which the line and continuum emission originate
from different regions in the galaxy, we caution that the spectra will
be biased toward the line-emitting regions. As the continuum emission
will be downweighted for these galaxies, the emission-line equivalent
widths may be overestimated. To check whether the emission line fluxes
may have been affected by the optimal extraction, we have repeated our
line fitting procedure for the boxcar extractions. We find no
systematic offset in emission line fluxes between the two
extractions. Finally, emission line measurements using the boxcar
extractions do not give a better agreement with the 3D-HST line
fluxes, thus suggesting that no significant bias has been intruced by
the optimal extraction method.

\subsection{Continuum Emission Sensitivities}\label{sec:cont}

\begin{figure} 
  \begin{center}           
     \includegraphics[width=0.48\textwidth]{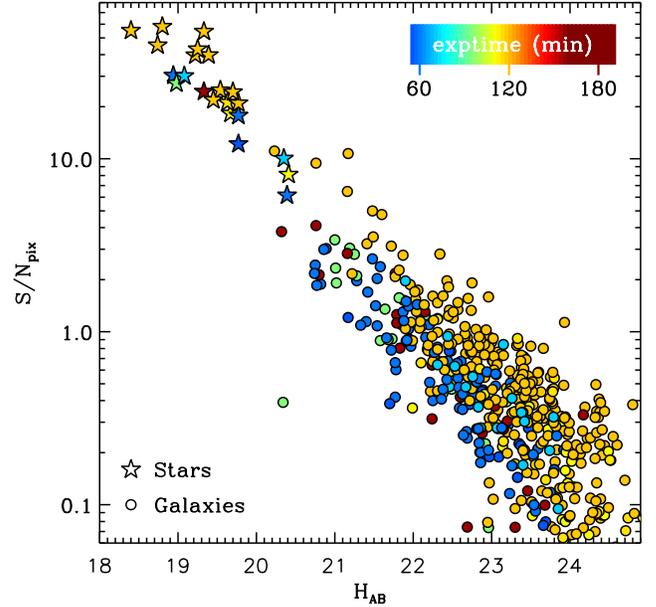} 

     \caption{The median S/N per pixel in the optimally extracted
       H-band spectra vs. the total H-band magnitude for all observed
       stars (stars) and galaxies (circles).  The color coding
       reflects the integration time. Serendipitous detections are not
       included in this figure. This figure does include bad weather
       masks, explaining the low S/N for galaxies with long
       integration times. \label{fig:cont}}   

  \end{center}                            
\end{figure}

In Figure~\ref{fig:cont} we show the S/N per pixel of the continuum
emission in the optimally extracted H-band spectra as a function of
the total F160W (H$_{\rm AB}$) magnitude. We take the median S/N level
in the wavelength interval covered by all spectra, given in
Section~\ref{sec:noise}. We show both the measurements for the
galaxies and the slit stars of all masks. The symbols are color coded
by the total exposure time. For galaxies at $z\sim2.3$, one pixel in
the H-band corresponds to 0.44~\AA\ in rest-frame, and thus the S/N
per rest-frame \AA\ is $\sim$1.5 times larger.

There is a large range in sensitivities among the different continuum
detections for the same integration time and F160W magnitude. This
scatter reflects the range in weather conditions, and the difference
in structural properties of the galaxies. For example, for more
extended galaxies or for larger values of the seeing, both the extraction
aperture and the slit losses are larger, resulting in lower values for
the S/N. 

To measure the S/N in each filter, and directly compare the depth and
seeing conditions of the different masks, we calculate the 3$\sigma$
depth (per pixel) for a point source directly from the spectra. For
this measurement we optimally extract a noise spectrum using the
profile of the slit star on several empty areas on the detector. Next,
we take the median noise level over the wavelength region targeted by
all spectra (see Section~\ref{sec:noise}). We multiply this value by a
factor of 3 to derive the 3$\sigma$ depth in $F_\lambda$. We convert
this depth to AB magnitude, using the effective wavelength of the
corresponding filter. As these depths are derived from the calibrated
data, both the AB magnitude and the $F_\lambda$ value are ``total'',
thus corrected for slit losses. 

\begin{figure} 
  \begin{center}           
     \includegraphics[width=0.49\textwidth]{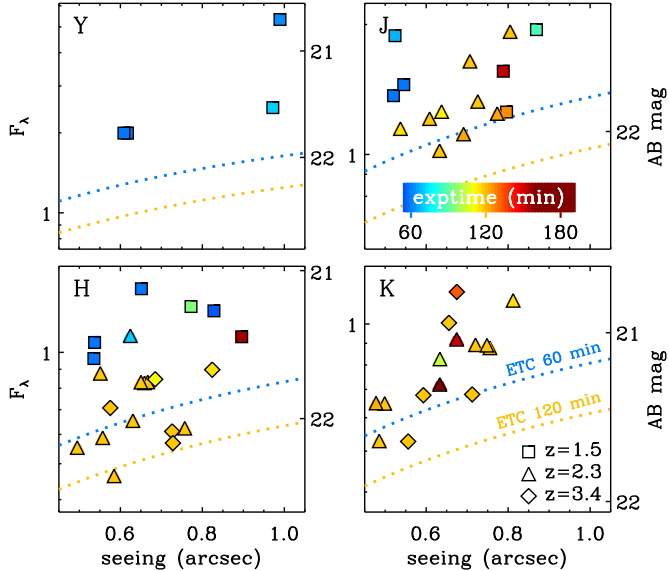} 

     \caption{The total flux $F_\lambda$ (in $10^{-18}$ erg s$^{-1}$
       cm$^{-2}$ \AA$^{-1}$) or AB magnitude of an artificial object
       for which we obtain a S/N of 3 per spectral pixel, as a
       function of seeing. Each data point represent a different mask
       and filter, the colors reflect the exposure time, and the
       symbols indicate the targeted redshift of the mask. The depths
       are derived from the optimally extracted 1D spectra in empty
       parts of the detector using the slit star profile, and thus are
       only valid for faint point sources. For a fair comparison, we
       take the median noise level in the wavelength regime covered by
       all spectra. 3\,$\sigma$ depths as derived by the exposure time
       calculator (ETC) for 60 and 120 min using the same method and
       wavelength range are presented by the dashed blue and orange
       lines, respectively. \label{fig:sens}}   

  \end{center}                            
\end{figure}

Figure~\ref{fig:sens} presents an overview of the depth for all 21
masks as a function of the effective seeing of the reduced
spectra. Each panel represents a  different filter, and each symbol
represents one mask for a specific filter. The symbols are color coded
by integration time. The sensitivities in Figure~\ref{fig:sens} are
only valid for faint point sources. The noise level increases, and
thus the depth decreases, for extended sources, due to larger slit
losses and a larger extraction aperture. For very bright sources the
depth will also decrease, as the noise is not merely determined by the
background level, and the source itself contributes to the
noise level as well.

Figure~\ref{fig:sens} also shows the median 3$\sigma$ depth given by
the official MOSFIRE exposure time calculator (ETC, by G. Rudie) using
the same wavelength region. We show the ETC results for a range in
seeing and 2 exposure times (one and two hours). The ETC does not
account for slit losses, and therefore, as input magnitude we have to
give the magnitude of the flux that falls in the extraction
aperture. Thus, we first convert the total magnitude to the extraction
aperture magnitude. For the extraction aperture we take 0\farcs 7
(slit width) times twice the FWHM of the effective seeing (see
Section~\ref{sec:ext}). We also correct the ETC S/N for an optimal
extraction, by dividing by a factor of 0.81. We derive this factor by
comparing the S/N of spectra extracted using an optimal extraction and
a boxcar extraction with the same extraction aperture. 

Figure~\ref{fig:sens} shows that the exposure time calculator is on
average optimistic by a factor of $\sim$2, though 2 out of 21 masks
in the H-band are consistent with the theoretical expectations for
their seeing (co2\_03 and co3\_04). Non-optimal or variable weather
conditions may contribute to the difference between the expected
and measured performance.

\begin{figure*}
  \includegraphics[width=0.33\textwidth]{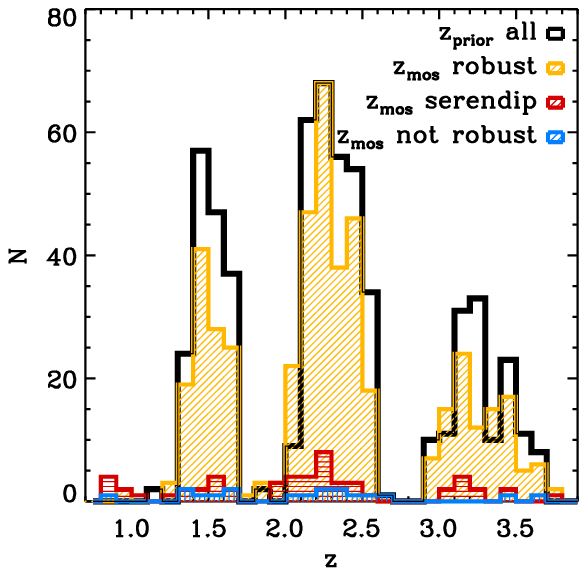}    
  \includegraphics[width=0.33\textwidth]{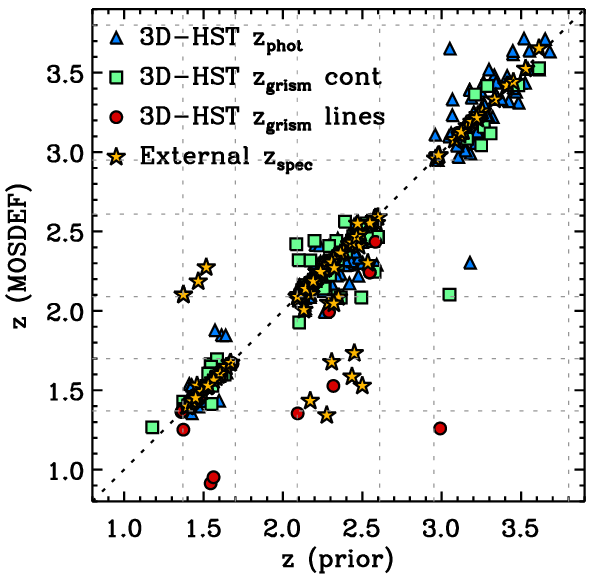}     
  \includegraphics[width=0.33\textwidth]{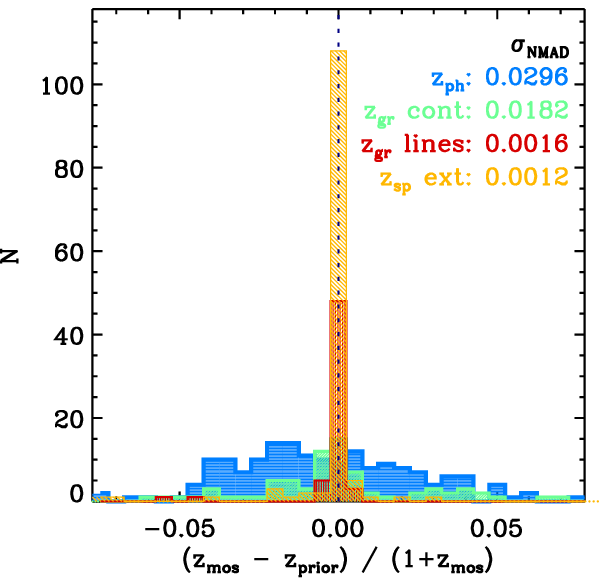}     

  \caption{Left: MOSDEF redshift distribution compared to the prior
    redshift distribution used for our target selection (black
    histogram). The MOSDEF redshift distribution is divided into
    robust redshifts (yellow histogram), inconclusive redshifts (blue
    histogram), and robust redshifts derived for serendipitous
    detections (red histogram). The latter galaxies are not included
    in the black histogram. Middle: Comparison of MOSDEF and prior
    redshifts for primary targets with robust MOSDEF redshifts. The
    prior redshifts consists of spectroscopic redshifts when available
    (yellow stars), 3D-HST grism redshifts, or 3D-HST photometric
    redshifts when no grism spectrum is available (blue
    triangles). Grism redshifts are divided between those which are based
    on emission lines (red circles) and those for which only continuum
    emission is detected (green squares). The gray dotted lines
    indicate the target redshift intervals. Right: Distribution of the
    difference in prior and mosdef redshifts for the four prior
    redshift classes, with the colors of the histograms corresponding
    to the colors of the symbols in the middle panel. The normalized
    median absolute deviation between the prior and mosdef redshifts are
    given for each prior redshift class.
    \label{fig:redshifts}}

\end{figure*}

\subsection{Absorption Line Redshifts}

For galaxies with strong continuum emission but no line emission, we
derive spectroscopic redshifts from absorption lines. There are 14
targets with a S/N$_{\rm pix} > 3$ in the H-band (see
Figure~\ref{fig:cont}). 8 out of 14 targets have detected emission
lines. Most of these galaxies also show clear absorption lines in
their spectrum as well. The six remaining galaxies all have quiescent
SEDs (as identified using their rest-frame $U-V$ and $V-J$ color, see
Section~\ref{sec:samplechar}, explaining the lack of detected emission
lines.

\begin{table}
\centering
\caption{Success rate of targeted galaxies}
\begin{tabular}{l | r r r | r r r | r | r}
\hline
Mask & \multicolumn{3}{c|}{$N_{\rm targeted}$} & \multicolumn{3}{c|}{$N_{\rm confirmed}$} & $F\footnote{Fraction of targeted galaxies for which we measure a robust spectroscopic redshift}$ & $N_{\rm ser}$\\
& $z_{\rm low}$ & $z_{\rm mid}$ & $z_{\rm high}$ & $z_{\rm low}$ & $z_{\rm mid}$ & $z_{\rm high}$ & \% & \\
\hline\hline
 ae1\_01 &    25 &     3 &     0 &    20 &     3 &     0 &    82 &     4  \\
 ae1\_05 &    25 &     2 &     1 &    15 &     2 &     0 &    61 &     3  \\
 co1\_03 &    26 &     2 &     1 &    16 &     2 &     0 &    62 &     0  \\
 co1\_05 &    20 &     5 &     1 &    10 &     2 &     0 &    46 &     3  \\
 gn1\_04 &    24 &     3 &     0 &    17 &     1 &     0 &    67 &     2  \\
 ud1\_01 &    32 &     1 &     0 &    29 &     1 &     0 &    91 &     0  \\
 ae2\_03 &     3 &    24 &     0 &     4 &    21 &     0 &    93 &     4  \\
 ae2\_04 &     0 &    27 &     2 &     1 &    25 &     1 &    93 &     2  \\
 ae2\_05 &     1 &    26 &     2 &     0 &    21 &     2 &    79 &     2  \\
 co2\_01 &     2 &    29 &     0 &     1 &    25 &     0 &    84 &     5  \\
 co2\_03 &     1 &    24 &     0 &     0 &    21 &     0 &    84 &     7  \\
 co2\_04 &     2 &    25 &     0 &     1 &    22 &     0 &    85 &     4  \\
 gn2\_04 &     1 &    24 &     2 &     1 &    19 &     1 &    78 &     2  \\
 gn2\_05 &     0 &    22 &     6 &     0 &    21 &     5 &    93 &     4  \\
 gn2\_06 &     1 &    24 &     3 &     2 &    17 &     2 &    75 &     1  \\
 gs2\_01 &     0 &    26 &     0 &     0 &    21 &     0 &    81 &     0  \\
 ae3\_04 &     3 &     2 &    26 &     2 &     2 &    22 &    84 &     0  \\
 co3\_01 &     0 &     8 &    21 &     0 &     9 &    11 &    69 &     3  \\
 co3\_04 &     0 &     3 &    26 &     1 &     3 &    18 &    76 &     0  \\
 co3\_05 &     3 &     2 &    25 &     2 &     2 &    24 &    93 &     6  \\
 gn3\_06 &     1 &     2 &    21 &     0 &     2 &    17 &    79 &     3  \\
\hline
   Total &   170 &   284 &   137 &   122 &   242 &   103 &    79 &    55  \\

\hline\hline
\end{tabular}
\tablecomments{ae1\_05, co1\_03 and co1\_05 are bad weather masks.}
\label{tab:success}
\end{table}

In order to determine a spectroscopic redshift, we fit the spectra (in
combination with the photometry) of the six targets without emission
lines by stellar population models, using the fitting code FAST
\citep{MKriek2009b}. For three out of six spectra, the fitting yields
robust spectroscopic redshifts, and for one spectrum the spectroscopic
redshift is less robust. One spectrum has no coverage in the
4000\,\AA\ break region, in which nearly all strong absorption lines
are expected. Thus, for this galaxy no robust redshift is
obtained. The remaining galaxy is too noisy to yield a spectroscopic
redshift. Two of the three spectra for which we measure a robust
redshift target the same galaxy (COSMOS-11982). The spectrum in mask
co2\_03 has a S/N of 9.4 per pixel in the H-band and the spectrum in
mask co3\_01 has a S/N of 4.1. By fitting the spectra independently,
we find the same spectroscopic redshift of $z=2.089$.

\subsection{Spectroscopic Success Rate}

An important factor for the overall success rate of our survey is the
fraction of galaxies for which we measure a robust spectroscopic
redshift. For 462 out of 591 primary MOSDEF galaxies we securely
identify and measure emission lines in the 1D extracted spectra.  In
cases for which only one emission line is significantly detected, the
spectroscopic redshift is classified as robust if it is consistent
with the photometric redshift within the typical photometric redshift
uncertainty. There are three additional spectra for which we securely
identify a redshift from multiple absorption lines. Thus, our
spectroscopic success rate is 79\% (465 out of 591). 31 of the 465
galaxies have been observed and confirmed in 2 masks, and thus the
number of unique galaxies with robust spectroscopic redshifts is 434.

Out of the 591 primary galaxies, 68 host an AGN, based on either their
IRAC colors or X-ray luminosity \citep{ACoil2015}. Thus, on average we
target 3.2 AGNs per mask. This number does not include AGNs that are
identified based on just their optical spectra. The spectroscopic
success rate for AGNs is 75\%. Three of the 51 confirmed galaxies
hosting an AGN have been observed twice.

Table~\ref{tab:success} gives an overview of the number of galaxies
targeted per redshift regime for each mask, and the number of galaxies
for each redshift regime for which we measure a robust redshift. Each
mask has a few fillers from the other redshift intervals. For nearly
all galaxies the prior redshifts fall in the targeted redshift
intervals\footnote{Due to catalog updates a few of the target
  redshifts scattered outside the targeted redshift intervals after
  the galaxies were observed with MOSFIRE.}. However, the MOSDEF
redshifts are in some cases outside the redshift intervals. Thus, when
giving the number of confirmed galaxies we use broader redshift
intervals (low: $z\le1.9$, middle: $1.9<z\le2.75$, high: $z>2.75$) in
Table~\ref{tab:success}. For each mask we give the spectroscopic
success rate, which is the ratio of all confirmed by all targeted
galaxies. The three bad weather masks have a lower success rate, with
an average of 55\%. If we do not include the bad weather masks, our
spectroscopic success rate is 82\%. 

There are 20 additional galaxies for which we detect emission or
absorption lines, but the spectroscopic redshifts are inconclusive due
to low S/N or the multiple emission lines yield inconsistent
redshifts. In addition, we have 55 robust redshifts for serendipitous
detections. The left panel of Figure~\ref{fig:redshifts} shows the
redshift distributions of the prior, robust, serendipitous and
non-robust redshifts. Most serendipitous detections fall in or near
the targeted redshift ranges. This is not surprising, as we are less
sensitive to picking up features in between the atmospheric windows
(see Figure~\ref{fig:features}). There are several serendipitous
detections, though, which have lower redshifts. 

In the middle and right panels of Figure~\ref{fig:redshifts} we
compare robust MOSDEF redshifts (of primary targets only) with the
prior redshifts used for target selection. The prior redshifts are a
combination of spectroscopic redshifts from primarily optical
spectroscopic surveys (see Section~\ref{sec:surdes}), 3D-HST grism
redshifts (including and excluding emission lines), and photometric
redshifts as derived from the 3D-HST photometric catalogs using
EAzY. For the majority of galaxies with prior spectroscopic redshifts
or with emission lines in the grism spectra we confirm the redshift,
with a normalized median absolute deviation \citep[$\sigma_{\rm
    NMAD}$,][]{GBrammer2008} of 0.0012 and 0.0016,
respectively. For most galaxies with photometric or grism redshifts
without detected emission lines, the MOSDEF redshifts are close to the
prior redshifts with a $\sigma_{\rm NMAD}$ of 0.30 and 0.18,
respectively. There are 16 galaxies that scatter from one to the other
redshift window. Interestingly, nearly all catastrophic failures are
galaxies with prior spectroscopic redshifts. 

For 36\% of the confirmed MOSDEF galaxies we had a spectroscopic
redshift (including grism emission line redshifts) prior to the
survey. This fraction does not include galaxies with incorrect prior
spectroscopic redshift. Thus, for 64\% of the targets only photometric
redshifts, grism continuum redshifts, or incorrect spectroscopic
redshifts were previously available.

\begin{figure}
  \includegraphics[width=0.48\textwidth]{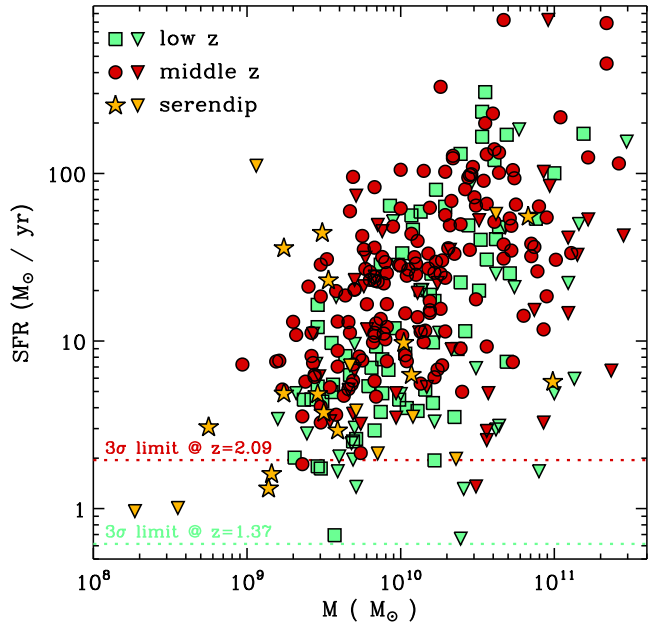}     

  \caption{SFR vs. stellar mass for galaxies in the low (green) and
    middle redshift interval (red) for a Chabrier IMF. 3$\sigma$ upper
    limits are indicated by the up-side-down triangles. The yellow
    colored symbols indicate the galaxies which were serendipitously
    detected. The SFR are derived from the combination of \ha\ and
    \hb. The 3$\sigma$ limits on the SFR calculated using the optimal
    3$\sigma$ line sensitivities in H and K, are shown for $z\sim1.37$
    and $z\sim2.09$ (i.e., the lower boundaries of the redshift
    intervals), respectively. In this calculation we assume no dust
    attenuation.
  \label{fig:sfr_mass}}

\end{figure}

\begin{figure*}[!t]
  \begin{center}           

     \includegraphics[width=0.49\textwidth]{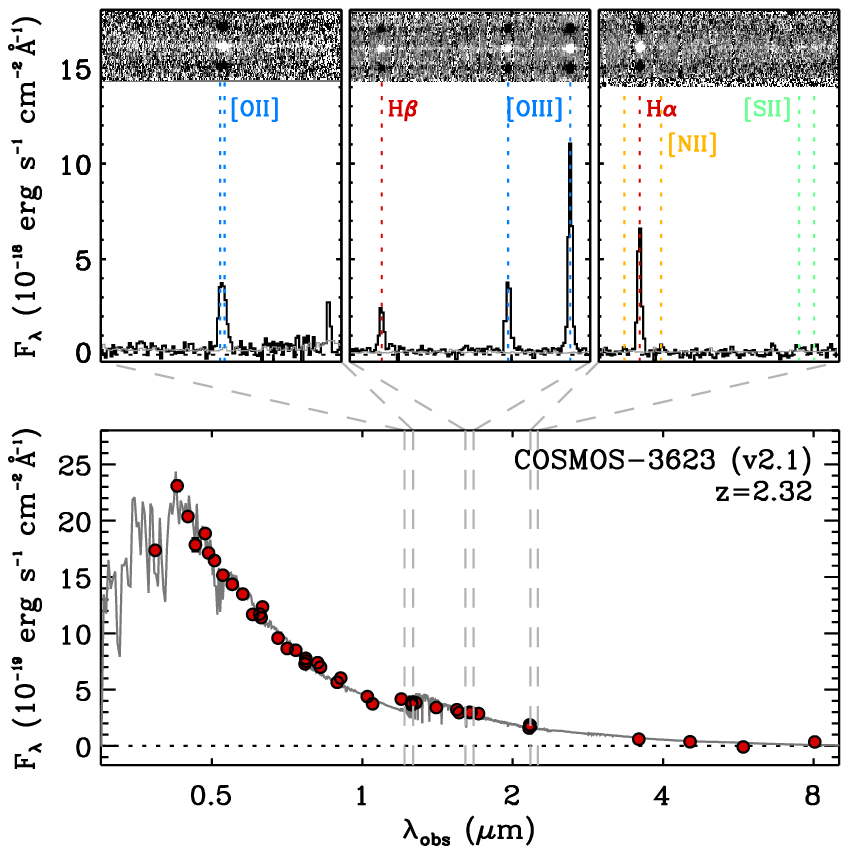}
     \includegraphics[width=0.49\textwidth]{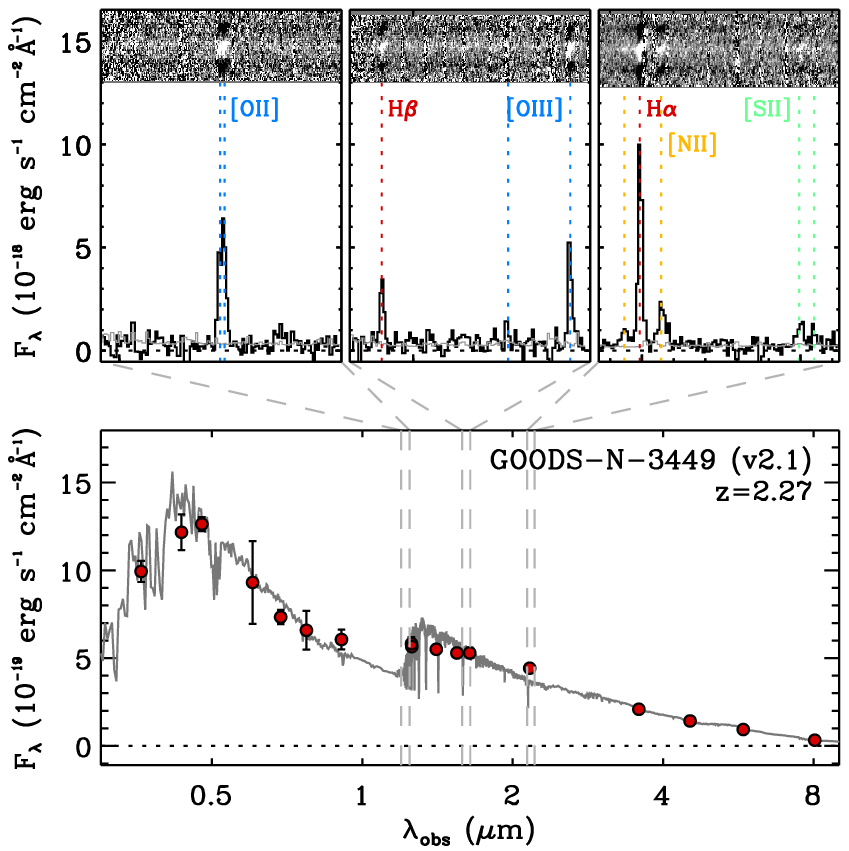}\\
     \includegraphics[width=0.49\textwidth]{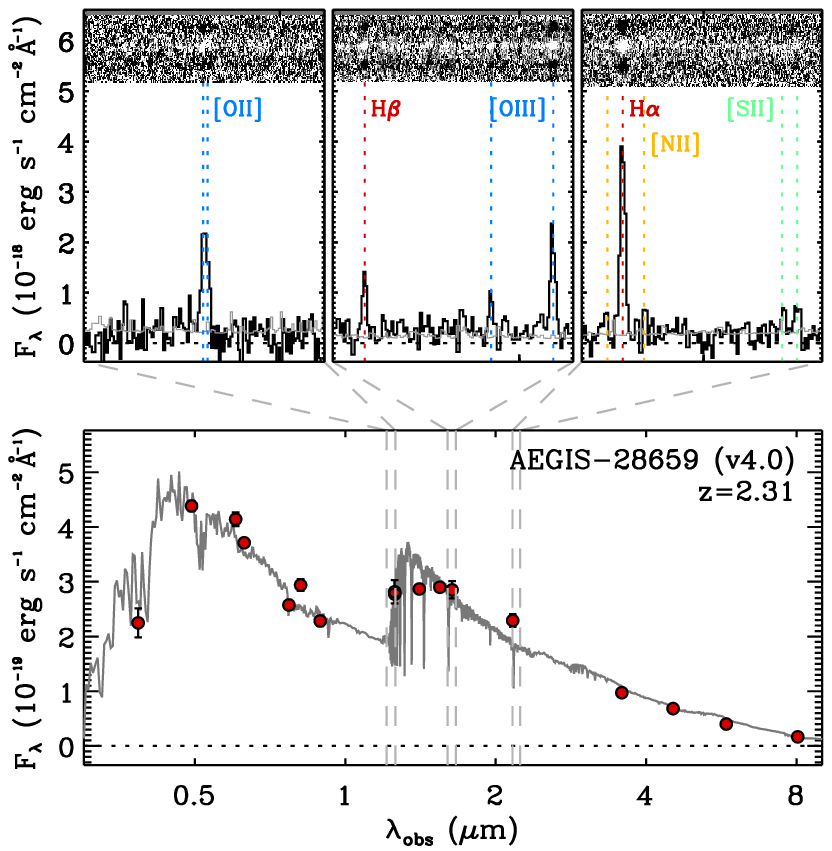}
     \includegraphics[width=0.49\textwidth]{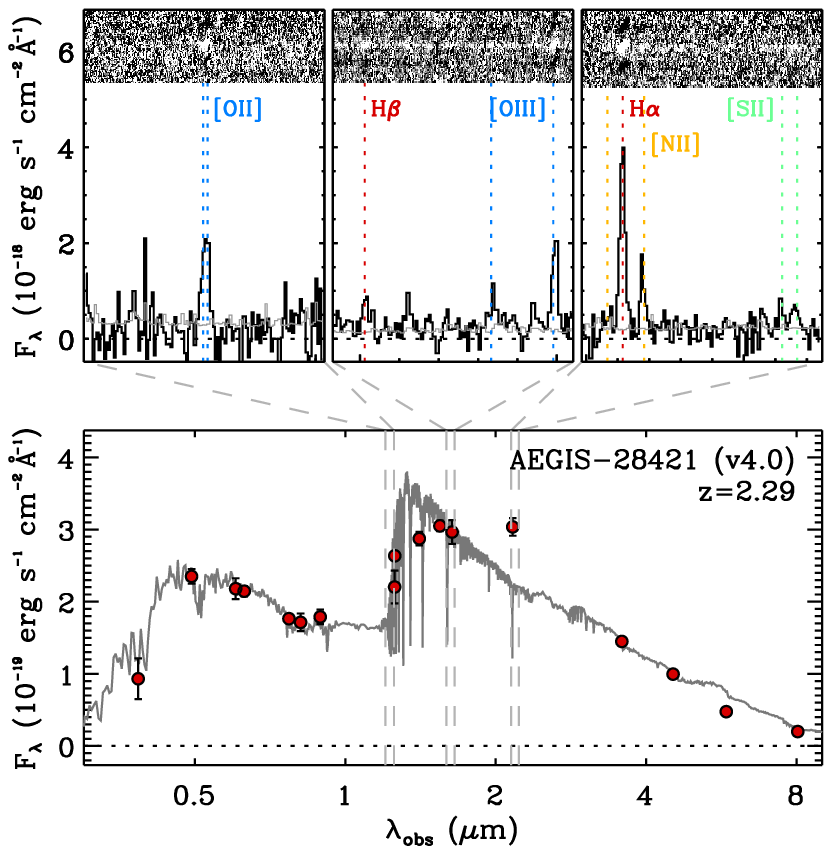}

     \caption{Example MOSFIRE spectra and corresponding
       multi-wavelength SEDs for 8 galaxies in the middle MOSDEF
       redshift regime. The galaxies have different SED shapes and are
       ordered by decreasing UV-to-optical flux. The lower panels show
       the rest-frame UV to near-IR photometry from the 3D-HST
       photometric catalogs \citep{RSkelton2014} and the best-fit
       stellar population model. The dashed vertical lines indicate
       the wavelength intervals for which we show the MOSFIRE spectra
       in the top panels. With the exception of COSMOS-11982, the top
       panels show both the 1D and 2D MOSFIRE spectra in the
       wavelength regions around the \oii\ doublet, \hb\ and \oiii,
       and \ha, \nii, and \sii, from left to right,
       respectively. Thus, we only show selected regions of the full
       MOSFIRE spectra. For COSMOS-11982 we zoom in around the
       absorption lines Ca\,{\sc ii} H and K in the J-band and Mgb in
       the H-band. All 1D spectra are binned by 3 pixels in wavelength
       direction, while excluding very noisy wavelengths (i.e.,
       corresponding to the locations of sky lines) and are shown in
       black. For COSMOS-11982 J-band we binned the spectra by 7
       pixels. The binned noise spectra are shown in gray. For clarity
       the 2D spectra have been stretched in the vertical direction by
       a factor of 2. \label{fig:seds}}

  \end{center}                            
\end{figure*}

\addtocounter{figure}{-1}

\begin{figure*}[!t]
  \begin{center}           
    \includegraphics[width=0.49\textwidth]{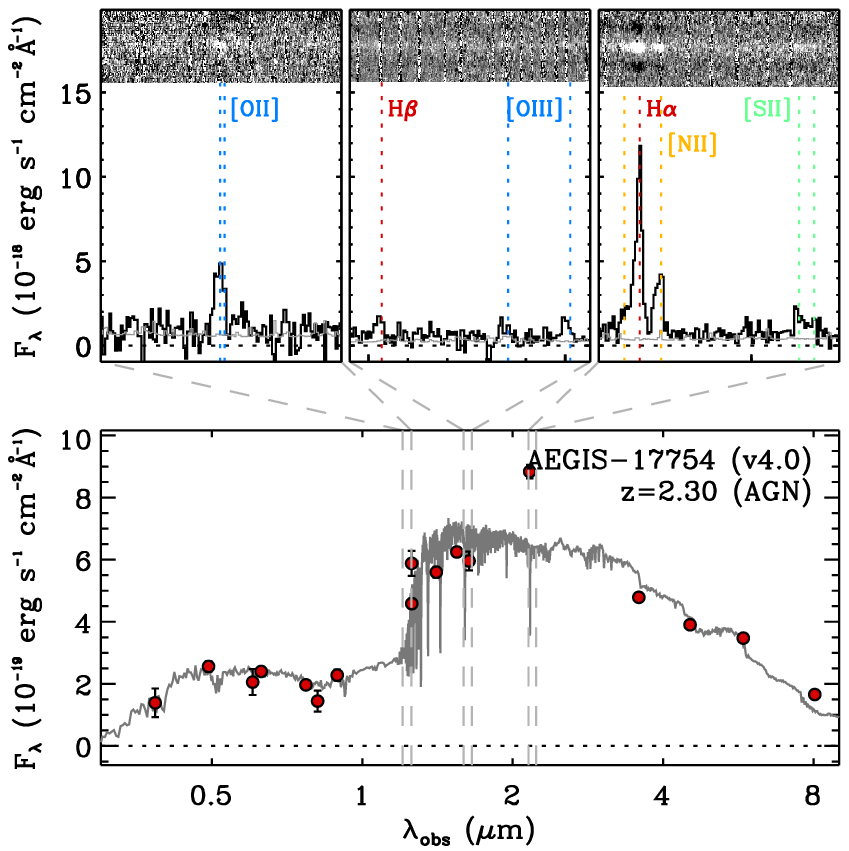}
    \includegraphics[width=0.49\textwidth]{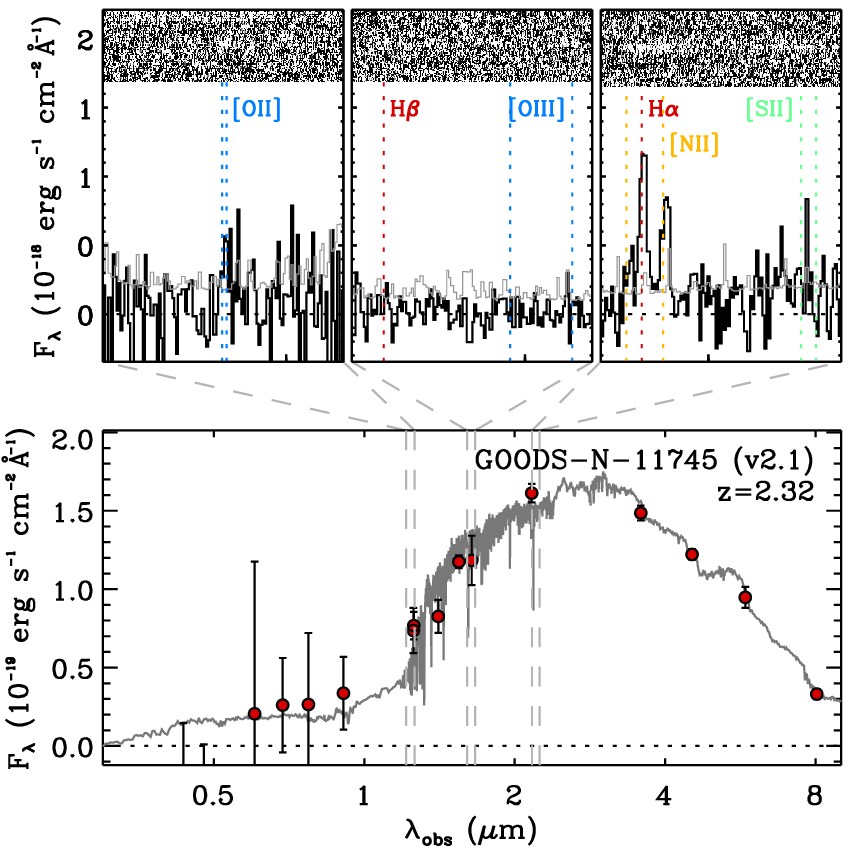}\\
    \includegraphics[width=0.49\textwidth]{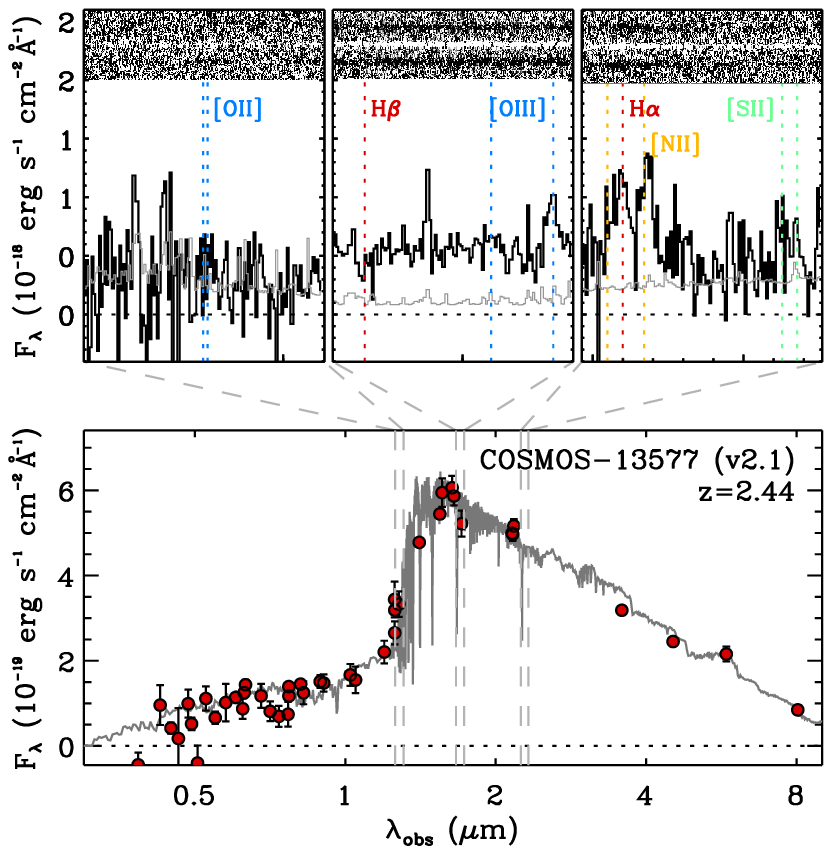}
    \includegraphics[width=0.49\textwidth]{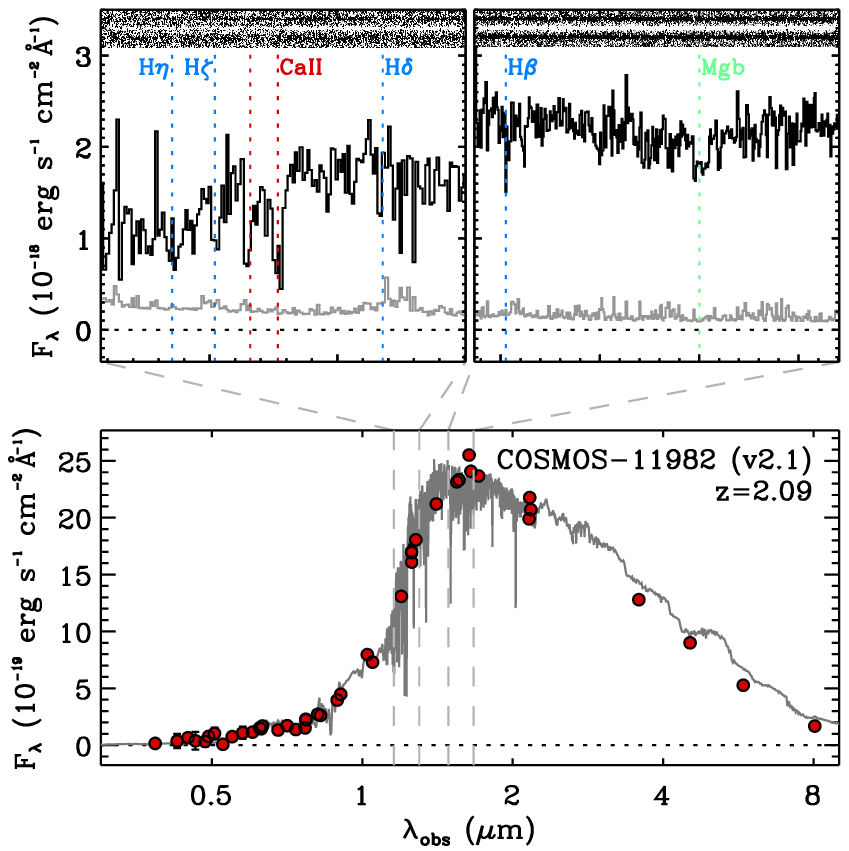}
    \caption{(Continued)}

  \end{center}                            
\end{figure*}

\subsection{Stellar Population Properties}\label{sec:starpop}

For all targeted galaxies as well as galaxies in the parent catalogs
we derive stellar population properties by comparing the photometric
SEDs with stellar population synthesis (SPS) models. The parent
catalogs are the trimmed versions of the 3D-HST catalogs by
\cite{RSkelton2014}, by imposing the redshift and magnitude criteria
for each redshift interval. We use the stellar population fitting code
FAST \citep{MKriek2009b}, in combination with the flexible SPS models
by \cite{CConroy2009}. We assume a delayed exponentially declining
star formation history of the form SFR $\propto t $ exp$(-t/\tau)$,
with $t$ the time since the onset of star formation, and $\tau$ the
characteristic star formation timescale. The age is allowed to vary
between $7.6<{\rm log} ~ (t/\rm yr) <10.1$ in steps of $\Delta({\rm
  log} ~ (t/\rm yr))=0.1$, but cannot exceed the age of the universe
at a given redshift. The star formation timescale $\tau$ can vary
between $8.0<{\rm log} ~ (\tau/\rm yr) <10.0$ in steps of $\Delta({\rm
  log} ~ (\tau/\rm yr))=0.2$. We furthermore assume a
\cite{GChabrier2003} stellar initial mass function (IMF) and the
\cite{DCalzetti2000} dust reddening curve. Spectroscopic redshifts from
MOSDEF are used when available. For galaxies without MOSDEF
redshifts, we use the prior redshift information. 

We include a template error function to account for template mismatch
in less constrained wavelength regions of the spectrum
\citep{GBrammer2008}. For example, the poorly understood thermally
pulsing asymptotic giant branch phase results in large uncertainties
in the rest-frame near-IR part of the spectrum, and thus the
wavelength range has a lower weight in the fit
\citep[e.g.,][]{CConroy2009,MKriek2010}. 1$\sigma$ confidence
intervals are derived using Monte Carlo simulations, by perturbing the
photometry using the photometric errors (corrected using the template
error function). Next, we determine the $\chi^2$ level that encloses
68\% of the simulations. We take the minimum and maximum values
allowed within this $\chi^2$ level as the confidence intervals on all
other properties \citep[see][for a more detailed
  description]{MKriek2009b}. While our default stellar population
parameters are derived using the method above, other MOSDEF papers may
use different methods \citep[Reddy et al. 2014,][]{ACoil2015}.

In addition to photometric SFRs, we also determine SFRs based on the
\ha\ and \hb\ emission lines. First, we derive a Balmer decrement from
the ratio of \ha\ to \hb. Both \ha\ and \hb\ are corrected for the
underlying Balmer absorption using the best-fit stellar population
model. By comparing this ratio to the intrinsic ratio of
\ha/\hb$=2.86$ for H\,{\sc ii} regions \citep{DOsterbrock1989} and
assuming the \cite{DCalzetti2000} attenuation curve, we derive the
reddening $E(B-V)$ and accordingly correct the H$\alpha$
luminosity. Finally, we convert the H$\alpha$ luminosity into a SFR
using the relation by \cite{RKennicutt1998}, adjusted for a
\cite{GChabrier2003} IMF (see Reddy et al. 2014 for a more detailed
description). In Figure~\ref{fig:sfr_mass} we show the SFRs and
stellar masses for the $z\sim1.5$ and $z\sim2.3$ MOSDEF galaxy
samples. This figure illustrates that our confirmed galaxies range in
stellar mass from $10^{9}-10^{11.5}~M_\odot$ and in SFR from $10^0-10^3~
M_\odot~ \rm yr^{-1}$.

In Figure~\ref{fig:seds} we show MOSDEF spectra, photometric SEDs, and
best-fit stellar population models for a variety of galaxies in the
middle redshift interval. The galaxies are ordered by decreasing
UV-to-optical flux ratio. COSMOS-3623 is a young and unobscured
star-forming galaxy with a strong Lyman break and a nearly absent
Balmer break. GOODS-N-3449 is slightly more evolved with a stronger
Balmer break. For AEGIS-28659 and AEGIS-28421 the UV gradually becomes
dimmer and the Balmer break becomes stronger. Both galaxies have
clearly detected emission lines. With AEGIS-17754, we continue the
sequence of a gradually increasing Balmer break. GOODS-N-11745 is a
very dusty star-forming galaxy. COSMOS-13577 and COSMOS-11982 both
have spectral energy distributions indicative of a quiescent stellar
population. COSMOS-13577 does have line emission, but the high ratio
of \nii/\ha\ indicates that it most likely originates from an AGN. For
COSMOS-11982 no emission lines are detected, and thus we zoom in on
the absorption lines.

\begin{figure*}[!hp]
  \begin{center}           

     \includegraphics[width=0.4\textwidth]{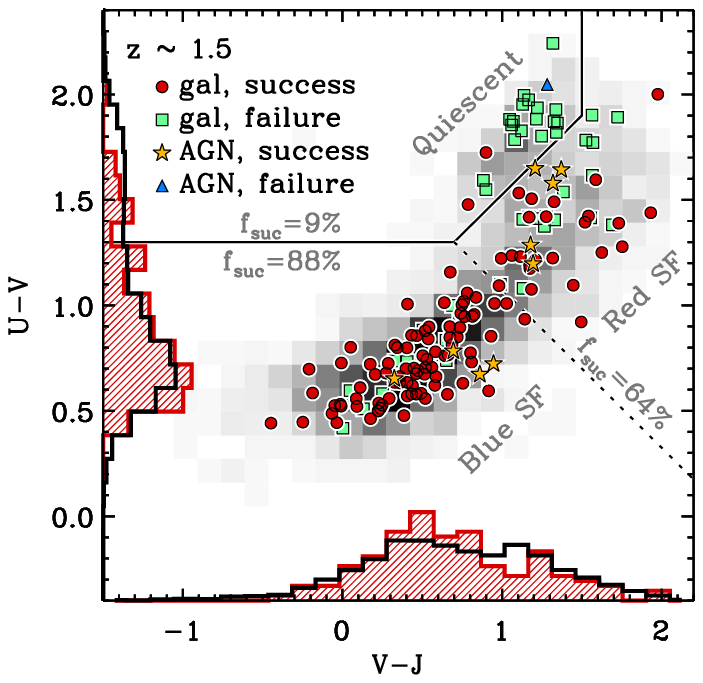}
     \includegraphics[width=0.4\textwidth]{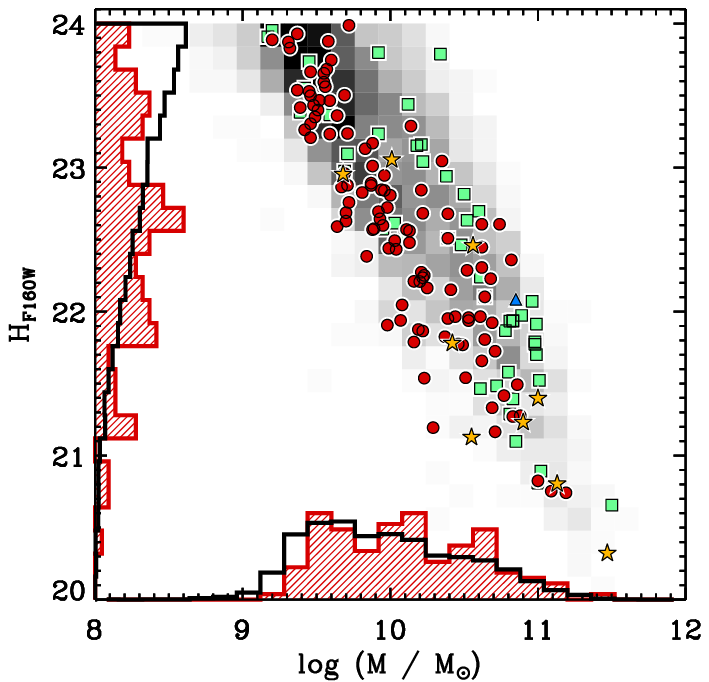}\\
     \includegraphics[width=0.4\textwidth]{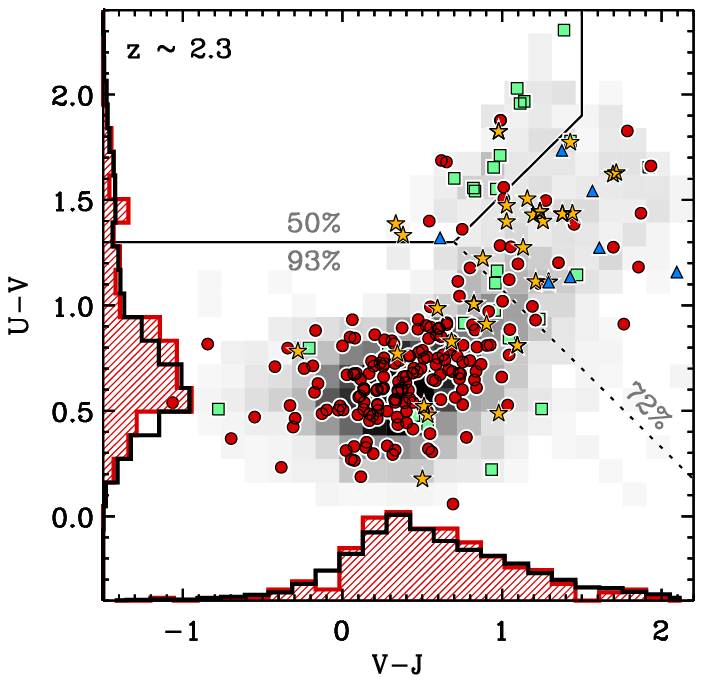}
     \includegraphics[width=0.4\textwidth]{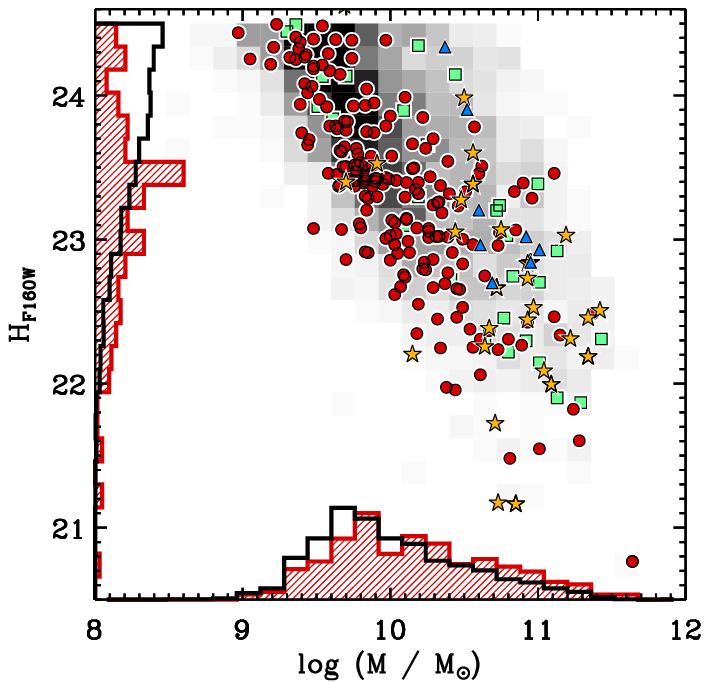}\\
     \includegraphics[width=0.4\textwidth]{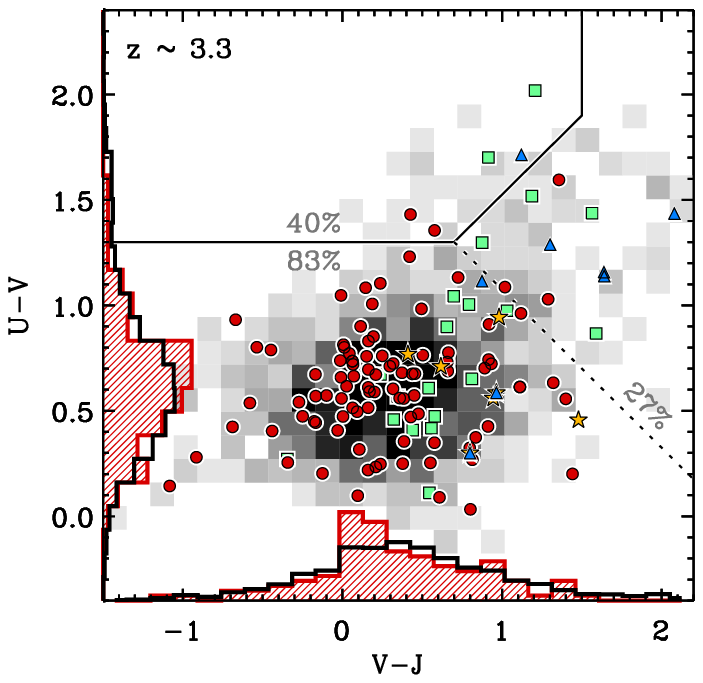}
     \includegraphics[width=0.4\textwidth]{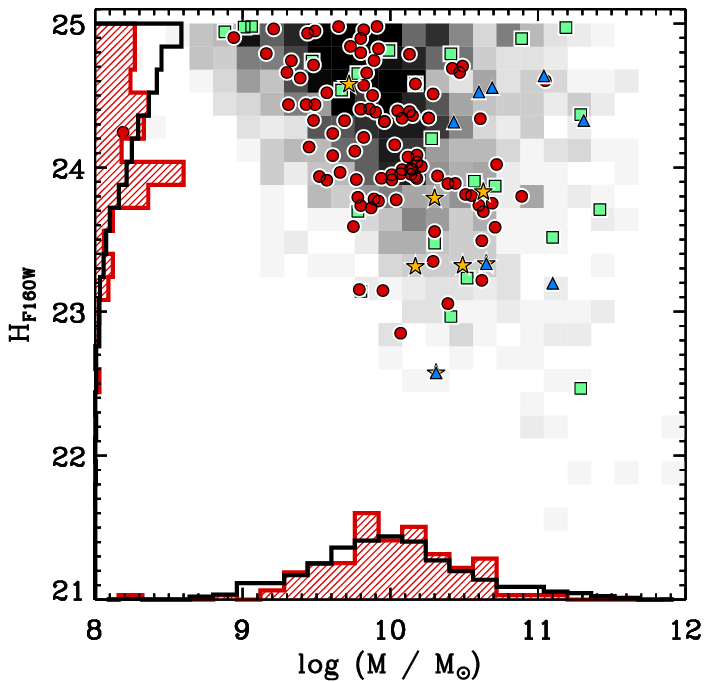}

     \caption{{\bf Left:} Rest-frame $U-V$ vs. $V-J$ color (UVJ
       diagram) for the three different redshift intervals. Quiescent
       galaxies populate the sequence enclosed by the box in the top
       left of this diagram. Star-forming galaxies populate the bottom
       sequence. In grayscale we show the parent galaxy samples
       from which the targets are selected, with H-band magnitude
       limits of 24.0, 24.5 and 25.0, for the low, middle, and high
       redshift intervals, respectively. The colored symbols represent
       our targeted objects, with red circles and yellow stars
       indicating galaxies and AGNs with robust spectroscopic redshifts. The
       green squares and blue triangles indicate targeted galaxies and
       AGNs for which no robust redshifts have been obtained. AGNs are
       identified by either their X-ray luminosity or IRAC colors
       \citep{ACoil2015}. The success rate for quiescent galaxies,
       blue star-forming galaxies (below/to the left of the dotted
       line) and red star-forming galaxies (above/to the right of the
       dotted line) are given in gray. The open black and shaded red
       histograms represent the distribution of the parent and
       confirmed samples (galaxies and AGN), respectively, for the
       property on the corresponding axis. The histograms are
       normalized to the same area. {\bf Right:} $H_{\rm F160W}$
       magnitude vs. stellar mass for the parent sample (grayscale)
       and the MOSDEF targeted galaxies and AGNs. Symbols and
       histograms are similar as in the left panels. \label{fig:uvj}}

  \end{center}                            
\end{figure*}

\subsection{Comparison to Full Galaxy Distribution} \label{sec:samplechar}

In the previous section we showed that our success rate is
high. Nonetheless, we are missing 21\% of the targeted galaxy
population. In this section we assess whether our targeted and
spectroscopically confirmed samples are representative of the full
galaxy sample in the same redshift interval to the same magnitude
limit, or whether we may be missing galaxies with specific
properties. For this assessment we consider as the full galaxy
population the parent sample within the targeted redshift regime down
to the same H-band magnitude limit from which our spectroscopic sample
was selected. 

We first compare galaxies in the rest-frame $U-V$ color versus
rest-frame $V-J$ color diagram (UVJ diagram). Galaxies out to
$z\sim2.5$ show a natural bimodality in this color-color space, and
both the star-forming and quiescent galaxies span tight sequences
\citep[e.g.,][]{SWuyts2007,RWilliams2009,KWhitaker2011}. Thus, this
diagram is used to isolate quiescent from star-forming galaxies and
classify galaxies out to $z\sim4$ \citep[e.g.,][]{AMuzzin2013a}.  The
quiescent sequence is primarily an age sequence, with galaxies
becoming redder in both colors with increasing age
\citep[e.g.,][]{KWhitaker2012,KWhitaker2013}. The star-forming
sequence primarily reflects the change in dust attenuation, with the
dustiest galaxies having the reddest $U-V$ and $V-J$ colors
\citep[e.g.,][]{GBrammer2011}. 

We derive rest-frame colors for all galaxies in the MOSDEF and parent
samples, using the EAzY code \citep{GBrammer2008} and following the
method described by \cite{GBrammer2009}. This method assumes a
redshift and interpolates between different observed bands using
templates that span the full range in galaxy properties. However, the
derived colors are not pure template colors, as the templates are only
used to fit the photometric data points closest to and surrounding the
specific rest-frame filter. When deriving rest-frame colors we assume
the MOSDEF or other spectroscopic redshifts when available. When no
spectroscopic redshift is available, we assume the best-fit grism or
photometric redshift.

In the left panels of Figure~\ref{fig:uvj} we show UVJ diagrams with
the parent distribution in grayscale and the targeted sample
presented by the colored symbols. Each row represents a different
redshift interval. Spectroscopically confirmed galaxies and AGNs are
indicated by red circles and yellow stars, respectively. As several
MOSDEF redshifts lie outside the targeted redshift interval, we adjust
the intervals for the confirmed galaxies to $1.25\le z\le 1.9$, $1.9 <
z \le 2.61$, and $2.94\le z \le 3.80$. Targeted galaxies and AGNs for
which no spectral features are detected, or for which the
spectroscopic redshifts are not robust, are presented by green boxes
and blue triangles, respectively. For these latter galaxies rest-frame
colors are derived assuming prior redshift or non-robust MOSDEF
redshift, respectively. The histograms show the distribution of $U-V$
and $V-J$ colors for the parent sample in black and the
spectroscopically confirmed sample (both galaxies and AGNs) in red. 

The location of galaxies in the UVJ diagram is dependent on mass, with
more massive galaxies populating the red sequence and the dusty part
of the star-forming sequence \citep[e.g.,][]{RWilliams2010}. In order
to further compare our sample to the parent population, we show
$H_{\rm F160W}$ magnitude versus stellar mass for the parent and
targeted galaxies in the right panels of Figure~\ref{fig:uvj}, using
the same symbols as in the left panels. For galaxies and AGNs for
which no spectroscopic features are detected, we derive stellar masses
assuming prior redshift information. The distributions in stellar mass
and $H_{\rm F160W}$ magnitude for the parent and spectroscopically
confirmed samples are presented by the black open and red shaded
histograms, respectively.

\begin{figure}
  \begin{center}           

     \includegraphics[width=0.48\textwidth]{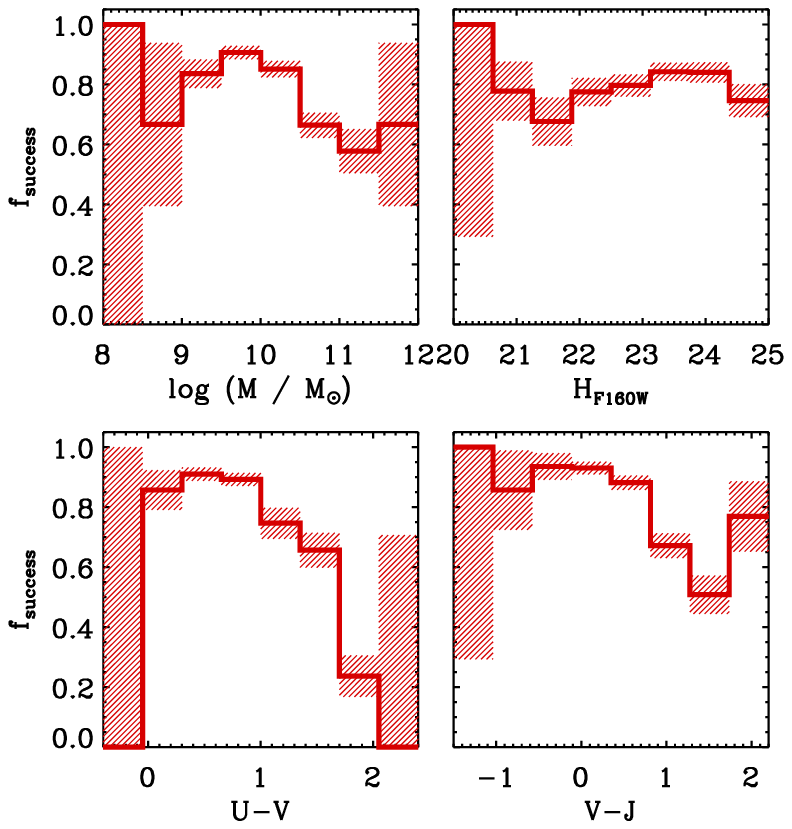}

     \caption{Success rate as a function of stellar mass (top left),
       H-band magnitude (top right), rest-frame $U-V$ color (bottom
       left) and rest-frame $V-J$ color (bottom right). The success
       rate is defined as the ratio of targeted galaxies with robust
       MOSDEF spectroscopic redshift to the total sample of targeted
       galaxies. The shaded regions indicate the uncertainties on the
       ratios. Serendipitous detections are excluded. The success rate
       primarily correlates with rest-frame $U-V$
       color. \label{fig:h_success}}

  \end{center}                            
\end{figure}

For all redshift ranges, our targeted sample is distributed over the
full UVJ diagram; we target galaxies along the entire star-forming and
quiescent sequences. The histograms show that the distribution of
rest-frame $U-V$ and $V-J$ colors of the parent and confirmed samples
are similar for the middle and high redshift interval. In the low
redshift interval we miss more galaxies with the reddest $U-V$ colors,
which we discuss in more detail in the next section. As in
the parent sample, the majority of the confirmed galaxies are blue in
both colors, and thus our sample is dominated by blue star-forming
galaxies. Red star-forming and quiescent galaxies form a minority in
our sample.

The small fraction of red star-forming and quiescent galaxies may
not be surprising, as these galaxies primarily populate the high-mass
end of the galaxy distribution, and thus will be sparse for a galaxy
sample with a mass limit of $\sim 10^9~M_\odot$. However, in our
selection scheme, we specifically aim to obtain a roughly flat
distribution in $H_{\rm F160W}$ magnitude and stellar mass, and thus
we prioritize galaxies by $H_{\rm F160W}$ magnitude. The right panels
in Figure~\ref{fig:uvj} indeed show that our sample is biased toward
brighter and slightly more massive galaxies compared to the parent
sample. The fact that our $H_{\rm F160W}$ prioritization did not
result in a bias toward redder rest-frame $U-V$ and $V-J$ colors is
due to the lower spectroscopic success rate of red galaxies.

\subsection{Success Rate for Different Galaxy Types}

In order to assess the spectroscopic success rate for different types
of galaxies, we split the UVJ diagram into three regions -- quiescent
galaxies, blue star-forming galaxies and red star-forming galaxies --
and give the success rate for each galaxy class in
Figure~\ref{fig:uvj}. For all redshift intervals the success rate is
highest ($\sim$90\%) for blue star-forming galaxies. Red star-forming
galaxies have a lower success rate, which varies from 27\% in the
highest redshift interval to 75\% in the middle redshift interval.
Quiescent galaxies have the lowest success rate, varying from 9\%
in the low redshift interval to 50\% in the middle redshift interval.

We also show the fraction of confirmed to targeted galaxies as a
function of stellar mass, H-band magnitude, rest-frame $U-V$ color and
rest-frame $V-J$ color in Figure~\ref{fig:h_success}. This figure
illustrates that success rate primarily correlates with rest-frame
$U-V$ color. The success rate also slightly decreases with increasing
stellar mass, which reflects the larger fraction of red
galaxies at the high mass end of the galaxy distribution. 

\begin{figure*}
  \begin{center}
     \includegraphics[width=0.4\textwidth]{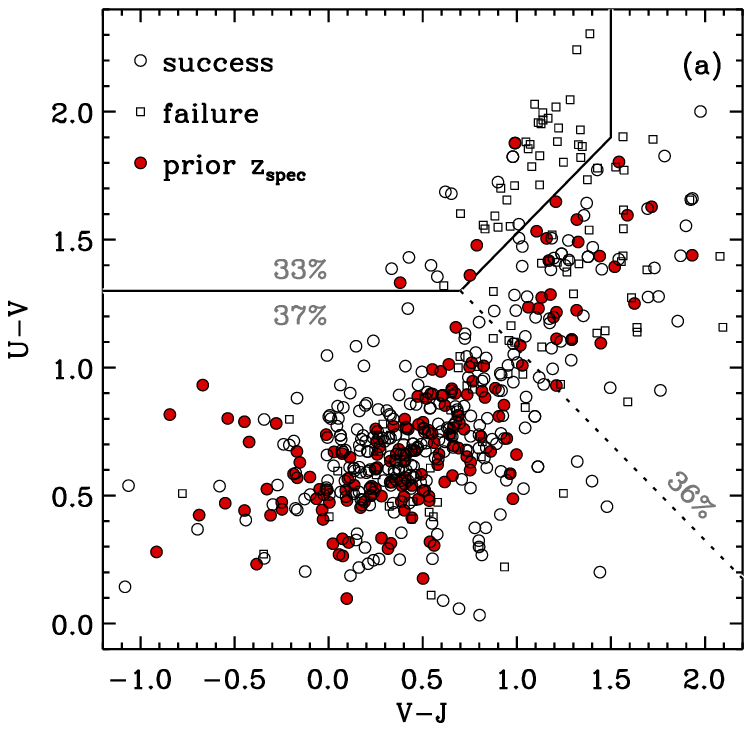}
     \includegraphics[width=0.4\textwidth]{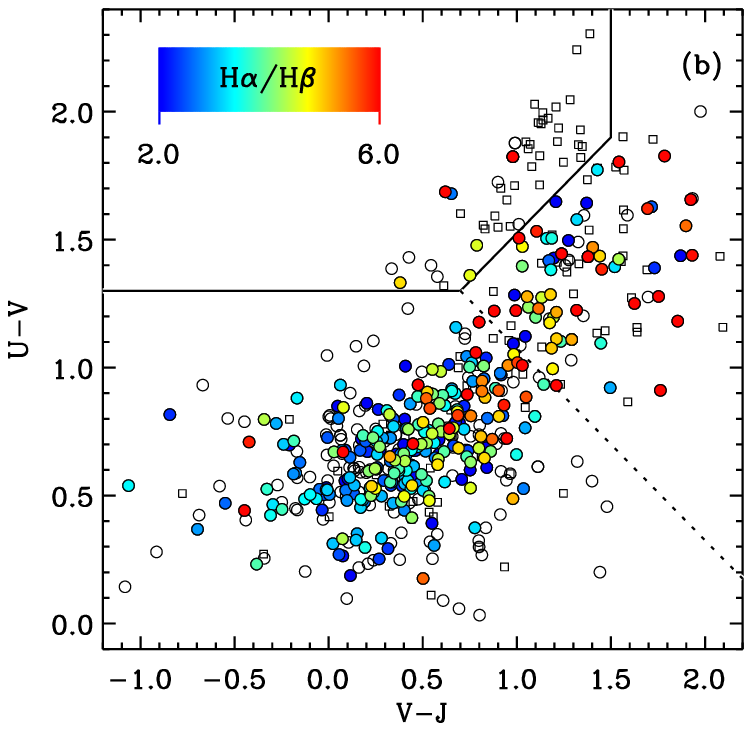}
     \includegraphics[width=0.4\textwidth]{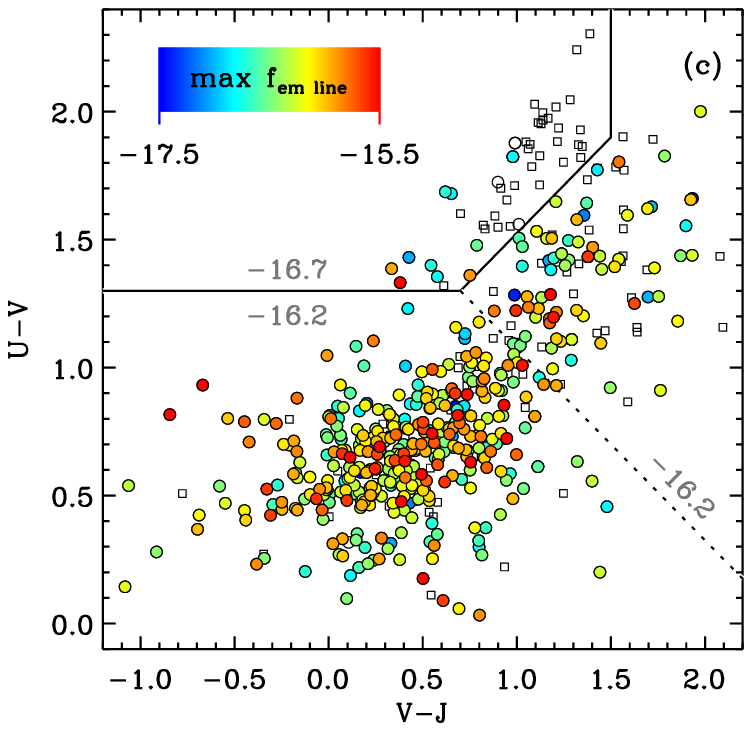}
     \includegraphics[width=0.4\textwidth]{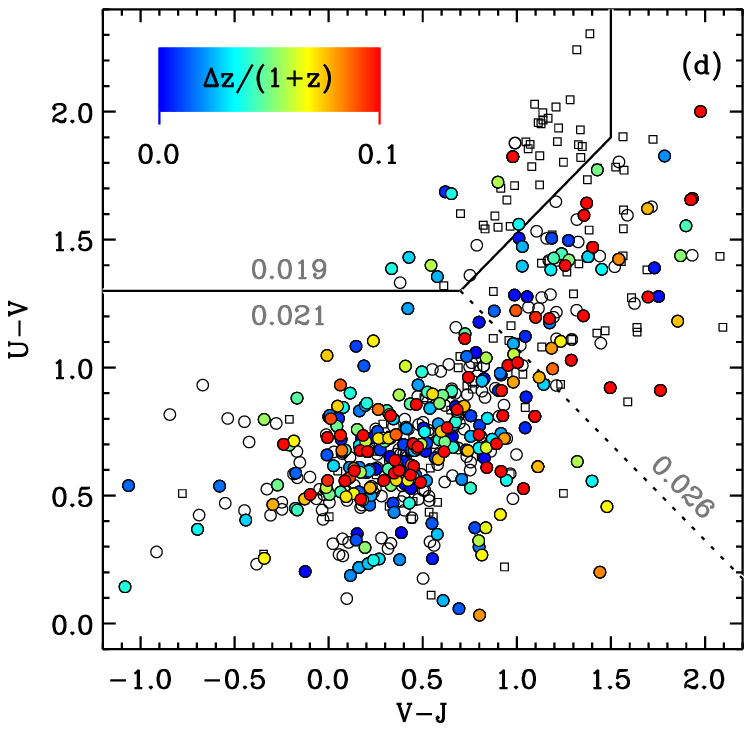}
     \includegraphics[width=0.4\textwidth]{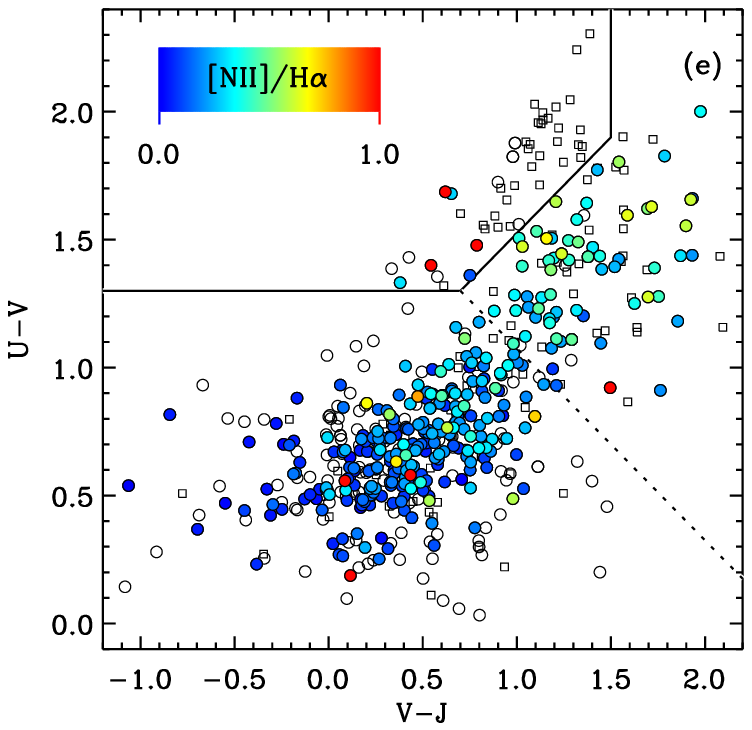}
     \includegraphics[width=0.4\textwidth]{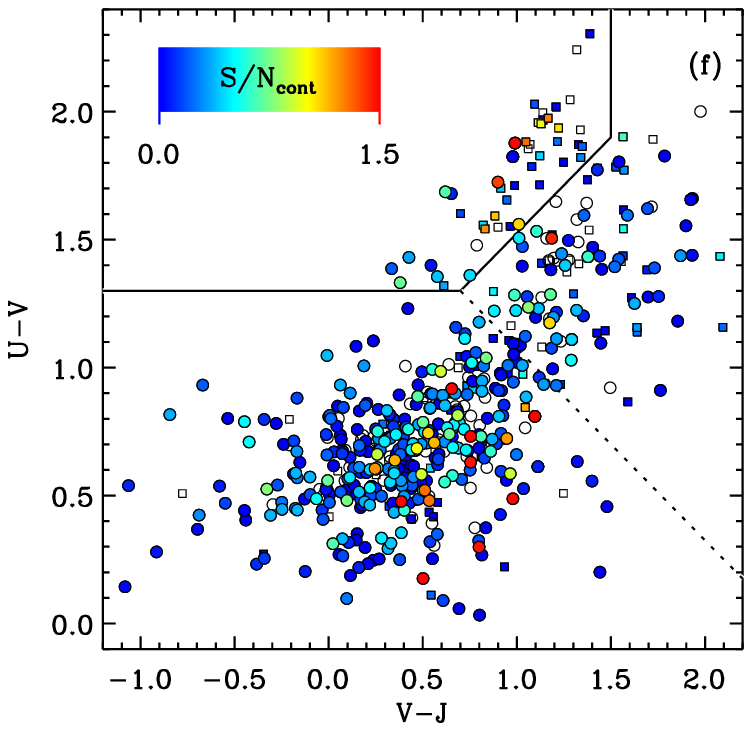}

     \caption{Assessment of spectroscopic success rate for different
       spectral types using the UVJ diagram. Galaxies with and without
       robust MOSDEF spectroscopic redshifts are indicated by circles
       and squares, respectively. (a)
       Galaxies with correct prior spectroscopic redshifts are
       indicated in red. The fraction of prior to MOSDEF spectroscopic
       redshifts (given in the panel) does not vary with spectral
       type. (b-c) Galaxies are color coded by the ratio of \ha/\hb,
       and the flux of the brightest emission line (log\,($f$/(erg
       s$^{-1}$ cm$^{-2}$))), when available. Red star-forming
       galaxies are more dusty, as indicated by their higher
       \ha/\hb\ ratio, but the median line emission (given in panel
       c), is the same for blue and red star-forming galaxies. (d)
       Galaxies with MOSDEF redshifts, but without prior spectroscopic
       redshifts are color coded by $\Delta z/(1+z)$, with the median
       value given for each class. Red star-forming galaxies have the
       least certain prior redshifts, but the difference is small. (e)
       Galaxies are color coded by the ratio of \nii/\ha, when
       available. This panel shows that for half of the quiescent
       galaxies for which we can measure this ratio, the line emission
       likely originates from an AGN. (f) All targeted galaxies are
       color coded by the median continuum S/N per pixel in the band
       which targets the 4000\,\AA\ break. The S/N for most quiescent
       galaxies is too low to measure absorption lines, or a spectrum
       is missing altogether.\label{fig:uvj2}}

  \end{center}
\end{figure*}

We examine the possible causes affecting the success
rate for the different galaxy types. We first assess whether our
prioritization scheme may affect the difference in the success rate
between the different classes. In our selection we prioritize galaxies
with prior spectroscopic redshifts. For these galaxies we are more
confident that the spectral lines will fall in observable parts of the
spectrum than for galaxies with just photometric
redshifts. Furthermore, the fact that these galaxies had prior
spectroscopic redshifts may suggest that they are bright or have
strong spectral features. Thus, if the prioritization by prior
spectroscopic redshift favors a particular galaxy type, it could lead
to a higher success rate for that class. In Figure \ref{fig:uvj2} we
show the UVJ diagram for all targeted galaxies, with the red circles
indicating the confirmed galaxies for which we had a correct prior
spectroscopic redshift. Interestingly, the fraction of galaxies with
prior spectroscopic redshifts compared to all confirmed galaxies does
not vary with galaxy types ($\sim$36\%). Hence, for all types we
increase the number of galaxies with spectroscopic redshifts by a
factor of $\sim$3. Thus, our prioritization by prior spectroscopic
redshift is not contributing to the higher success rate for blue
star-forming galaxies.

Another possible cause for the lower success rate of red galaxies may
be the decreasing strength of emission lines in redder galaxies. The
star-forming sequence is thought to be a sequence of increasing dust
attenuation, and redder galaxies may therefore have fainter emission
lines. To assess this theory, we show the UVJ diagram in
Figure~\ref{fig:uvj2}b color coded by the ratio of \ha/\hb. We find
that the \ha/\hb\ ratio is indeed higher for red star-forming
galaxies, indicative of more dust extinction (see Reddy et
al. 2014). However, to assess whether the larger dust extinction
results in lower line fluxes, we color code the UVJ diagram by the
flux of the strongest emission line in
Figure~\ref{fig:uvj2}c. Interestingly, the blue and red star-forming
galaxies have the same median flux of $10^{-16.2}\, \rm erg\, s^{-1}\,
cm^{-2}$. However, this finding does not rule out that we may miss
fainter emission lines for the redder star-forming population. We do
note that the average number of detected emission lines for blue
star-forming galaxies is higher than for red star-forming galaxies,
with 2.8 and 2.3, respectively.

Yet another possibility for the lower success rate of red star-forming
galaxies may be the more uncertain photometric redshifts compared to
those of blue star-forming galaxies. To test this scenario, in
Figure~\ref{fig:uvj2}d we color code all galaxies with a MOSFIRE
redshift, but without a prior spectroscopic redshift by $\Delta
z/(1+z)$, the difference in redshift between their prior and MOSDEF
redshift. With a median $\Delta z/(1+z)$ of 0.025, red star-forming
galaxies have the most uncertain prior redshifts, but the difference
among the different galaxy types is small. In order to
identify the primary reason for the lower success rate of red
star-forming galaxies, we would need emission line measurements and
spectroscopic redshifts for star-forming galaxies that have not been
confirmed.

For quiescent galaxies we find that the faint emission line fluxes are
contributing to the low success rate. Given that the galaxies in the
quiescent box have much lower SFRs, we indeed expect to find little or
no line emission. Nonetheless, for $z\sim2.3$, about half of the
galaxies in the quiescent box have detected emission lines. This
result is consistent with the results from \cite{MKriek2008b}, who
find that about 40\% of the quiescent galaxies at $z\sim2.3$ have line
emission. To test whether the line emission from our
quiescent galaxies originates from AGNs, we color code all galaxies in
the UVJ diagram by \nii/\ha\ ratio \citep[Figure~\ref{fig:uvj2}e, see
  als][]{ACoil2015}. For half of the galaxies in the quiescent box
with emission lines for which we can measure \nii/\ha, we indeed find
a high ratio ($>0.6$), indicative of an AGN. 

For quiescent galaxies we can also measure spectroscopic redshifts
from absorption lines. We only measured absorption line redshifts for
two $z\sim 2.3$ quiescent galaxies. To assess why this fraction is so
low, we color code the UVJ diagram by the continuum S/N per pixel in
the band that targets the 4000\,\AA\ break, as this wavelength region
covers several strong absorption lines. Figure~\ref{fig:uvj2}f shows that
most quiescent galaxies have a very low S/N in this wavelength region,
and 18\% of the quiescent galaxies (3 masks) are lacking data in this
wavelength region altogether. The galaxies with low S/N and missing
wavelength coverage are primarily in the low redshift masks. Three out
of the six low redshift masks were observed during bad weather
conditions, and the nominal integration time for this redshift
interval is shorter than for the higher redshift masks. The
integration times were shortened as in general galaxies are brighter
at lower redshift. However, as suggested by their location in the UVJ
diagram, quiescent galaxies at $z\sim1.5$ are likely older with lower
$M/L$, and thus may be more challenging to confirm
\citep{KWhitaker2012}. In addition, the fraction of quiescent galaxies
with emission lines is lower at $z\sim 1.5$. These factors together
may explain the very low success rate of the $z\sim 1.5$ quiescent
galaxies. 

In summary, the low success rate of quiescent galaxies is primarily
due to the low fraction of quiescent galaxies with detected line
emission and the low S/N or missing continuum emission around the
4000\,\AA\ break for most quiescent galaxies. Red star-forming
galaxies have brighter emission lines, and a higher fraction of
galaxies with detected line emission, and thus they have a higher
success rate than quiescent galaxies. However, the success rate for
red star-forming galaxies is lower than for blue star-forming
galaxies. The difference in success rates between these two galaxy
types is not well understood.

\section{SCIENCE OBJECTIVES OF MOSDEF}\label{sec:sci}

The MOSDEF dataset allows a wide range of new and unique studies,
which all contribute to constructing a complete picture of galaxy
formation. Most science cases rely on a combination of MOSDEF
rest-frame optical spectroscopy and other multi-wavelength datasets
available in the targeted fields. In this section we briefly summarize
our primary science objectives.

\subsection{Star Formation and the Mass Growth of Galaxies}

A fundamental aspect of the study of galaxy evolution is understanding
how galaxies build their stellar mass over cosmic time. By tracing
both the evolution in stellar mass and SFR of a complete galaxy
sample, we can constrain the rate and by which mechanism (star
formation vs.  mergers) galaxies grow
\citep[e.g.,][]{PvanDokkum2010,NReddy2012a}. However, deriving both
properties is challenging. In particular SFRs of $z>1$ galaxies are
highly uncertain due to the inaccessibility of reliable indicators
(\ha\ and \hb; bolometric flux).

With MOSDEF we detect -- for the first time -- both the \ha\ and
\hb\ emission lines for a large rest-frame optical magnitude-limited
sample of distant galaxies. These features together form the ideal SFR
indicator, which is relatively unbiased to dust extinction (unlike the
UV continuum emission). Using the multi-wavelength data
in the CANDELS fields, we will compare our SFRs with SFRs derived
using other indicators (e.g., UV, {\it MIPS} 24 $\mu$m, {\it
  Herschel}) to better calibrate SFR indicators and  obtain a full
census of star formation at high redshift. MOSDEF also improves the
accuracy of stellar mass measurements, by providing spectroscopic
redshifts and estimates of the contribution of line emission to the
photometric fluxes.

\subsection{Dust Attenuation}

A key aspect of quantifying SFRs is understanding how the intrinsic
galaxy spectrum is modulated by interstellar dust. Unfortunately, even
the deepest {\em Spitzer} and {\em Herschel} mid- to far-IR
observations are insufficient to directly detect dust emission from
individual $L^{\ast}$ galaxies at $z\ga 2$
\citep[e.g.,][]{NReddy2010,NReddy2012b}. Consequently, we are reliant
on stellar population modeling and the UV slope $\beta$, whose use for
estimating dust attenuation in particular types of galaxies has been
called into question by many studies
\citep[e.g.,][]{XKong2004,BSiana2009,NReddy2010,MKriek2013}.

MOSDEF will enable the measurement of one of the most direct and
locally well-studied dust indicators, the Balmer decrement
(H$\alpha$/H$\beta$), for a statistical sample of $1.4\lesssim
z\lesssim 2.6$ galaxies. Previous such measurements primarily relied on
stacked spectra \citep[e.g.,][]{ADominguez2013,SPrice2014}, or very
small samples of individual galaxies. Using the multi-wavelength data
in the MOSDEF survey fields, we will cross-check dust corrections
inferred from the Balmer decrement, UV slope, and (stacked) mid- and
far-IR emission, and build a complete census of dust properties of
$z\sim2$ galaxies (e.g., dust-to-star geometry, temperature). In a
first paper (Reddy et al. in prep) we present Balmer decrements for
$z\sim2$ star-forming galaxies and derive their dust attenuation
curves.

\subsection{Gas-Phase Metallicities}

The metal content of galaxies reflects the past integral of star
formation, modified by the effects of gas inflow (i.e., accretion) and
outflow (i.e., feedback). While the mass-metallicity relationship has
been measured for large samples of star-forming galaxies at $z<1$
\citep{CTremonti2004,JMoustakas2011}, observations of this trend at
higher redshifts have until recently been based either on small and/or
biased samples of individual objects \citep{FMannucci2009} or on
composite spectra that mask the variation among individual objects
\citep{DErb2006a}.

With MOSDEF we derive gas-phase metallicities for many individual
galaxies from a suite of bright rest-frame optical emission lines. In
a first paper \citep{RSanders2015} we correlate  gas-phase metallicity
with stellar mass and SFR of 86 star-forming galaxies, and show that
high-redshift galaxies do not fall on the local ``fundamental
metallicity relation'' \citep{FMannucci2009} among stellar mass,
metallicity, and SFR \citep[see also][]{CSteidel2014}.

\subsection{ISM Physical Conditions}

Measurements of multiple rest-frame optical emission lines are crucial
for understanding the physical conditions in the ISM of high-redshift
galaxies. At low redshift, star-forming galaxies follow a fairly
tight sequence in the space of \nii/\ha\ vs. \oiii/\hb, also known as
the BPT diagram \citep{JBaldwin1981}. However, small samples of
$z>1$ galaxies with measurements of \hb, \oiii, \ha, and \nii\ appear
to be systematically offset from the excitation sequence of
low-redshift galaxies \citep[e.g.,][]{AShapley2005}. The origin of
these differences may reflect fundamental differences in distant
H\,{\sc ii} regions \citep[e.g.,][]{XLiu2008,JBrinchmann2008} and may
have severe implications for metallicity measurements of galaxies.

MOSDEF significantly increases the number of distant galaxies with
\ha, \hb, \oiii, and \nii\ emission line measurements, such that we
accurately measure their location in the BPT diagram. Measurements of
\oii\ and \sii\ enable studies of additional ISM excitation
diagrams. In a first paper \citep{AShapley2015} we confirm the offset
of the excitation sequence and assess how the offset varies with
stellar mass, specific SFR, and SFR surface density of $z\sim2.3$ galaxies.

\subsection{Stellar feedback}\label{sec:baryons}

The process described as ``feedback" is considered a crucial component
in models of galaxy formation. Feedback commonly refers to large-scale
outflows of mass, metals, energy, and momentum from galaxies,
regulating the amount of gas available to form stars, as well as the
thermal properties and chemical enrichment of the intergalactic
medium. However, directly observing the inflow of gas into
galaxies -- especially during the epoch when they are assembling --
remains challenging.

Using nebular emission lines in the MOSDEF spectra in combination with
 existing and new UV spectroscopy\footnote{We are conducting a
  complementary observing campaign to obtain rest-frame UV
  spectroscopy for MOSDEF galaxies with DEIMOS}, we will measure the
speed of outflowing (blueshifted) or inflowing (redshifted) gas, and
correlate this speed with galaxy properties such as (specific) SFR, SFR
surface density, inclination, and size. In addition, the resolution
of MOSFIRE spectra will allow for detailed profile fitting of the
strongest rest-frame optical emission lines, which will highlight
deviations from symmetric Gaussian profiles and/or underlying broad
components, which may be indicative of extended, outflowing ionized gas
\citep{RGenzel2011,SNewman2012}.

\subsection{Nuclear Accretion and Galaxy Co-evolution}

Determining the causes and evolution of AGN triggering and fueling is
essential to understanding the formation and evolution of both black
holes and galaxies. Accretion onto supermassive black holes appears to
peak at a redshift of $z\sim 1 - 3$ \citep[e.g.,][]{GHasinger2005},
though the exact location of this peak and its dependence on black
hole mass and AGN luminosity are still unknown. The physical
relationship between AGN and their host galaxies at this key epoch is
also unclear. It has been difficult to make progress on these
questions due to the small number of AGNs at these epochs for which
rest-frame optical emission lines have been measured.

The MOSDEF spectra allow us to optically identify AGN using the BPT
diagram. We will complement the BPT diagram with X-ray and mid-IR
color selection criteria, and quantify the fraction of galaxies that
host an AGN over cosmic time.  We will also use the \oiii\ luminosity
of AGN to probe black hole accretion at $z\sim2$, and quantify the
connection between black hole and galaxy growth by relating the AGN
fraction and accretion rates to host galaxy properties.  In a first
paper \citep{ACoil2015} we test various optical AGN classification
diagnostics at $z\sim2$, including the BPT, MEx \citep{SJuneau2011},
and CEx diagrams \citep{RYan2011}.

\subsection{Dynamical Masses and Structural Evolution}\label{sec:dyn}

During the peak of star formation activity massive galaxies
($>10^{11}\,M_\odot$) show a wide diversity in galaxy properties, with the
population about equally divided between star-forming and quiescent
galaxies \citep[e.g., ][]{MKriek2008b, AMuzzin2013b}. These galaxies
are not simply the younger versions of elliptical and star-forming
galaxies today, but were smaller and denser at similar mass
\citep[e.g.,][]{RWilliams2010}. Both populations appear to grow inside
out, but the physical mechanism (e.g., minor mergers, progenitor
bias, cold streams and in situ star formation) responsible for the growth is
still subject to debate
\citep[e.g.,][]{JvandeSande2013,ANewman2012,MCarollo2013,ADutton2011,PvanDokkum2013}.

Identifying the dominant growth mechanism for both quiescent and
star-forming galaxies requires accurate mass, kinematic and size
measurements for a large and complete sample of distant galaxies. With
MOSDEF we measure velocity dispersions from rest-frame optical nebular
emission and stellar absorption lines. Combined with high-resolution
rest-frame optical imaging from CANDELS, we will measure dynamical masses
and study how both quiescent and star-forming galaxies grow in
size, velocity dispersion, and mass over cosmic time.

\section{SUMMARY}\label{sec:sum}

In this paper, we present the MOSDEF survey, a 47 night program with
MOSFIRE on the Keck I Telescope, to obtain intermediate-resolution
($R=3000-3650$) rest-frame optical spectra for $\sim$1500 galaxies at
$1.37\le z\le 3.80$. The survey is being executed in three
well-studied extragalactic legacy survey fields (AEGIS, COSMOS and
GOODS-N) and will cover $\sim$ 600 square arcmin. The galaxy sample is
split into three redshift intervals ($1.37\le z\le 1.70$, $2.09\le
z\le 2.61$, and $2.95\le z\le 3.80$), for which bright rest-frame
optical emission lines (\oii, \hb, \oiii, \ha, \nii, and \sii) fall in
atmospheric windows. Emphasis is given to the middle redshift
interval, which will contain half of our sample. The remaining
galaxies will be evenly split among the lower and higher redshift
intervals. The galaxies are selected using the multi-wavelength
photometric and spectroscopic catalogs from the 3D-HST survey down to
fixed H-band magnitude. The magnitude limits are $H=24.0$, $H=24.5$,
and $H=25.0$, for the low, middle, and high redshift interval,
respectively. Priority is given to brighter galaxies, galaxies with
more reliable redshifts, and galaxies hosting an AGN.

MOSDEF is scheduled to be executed over 4 spring semesters, and we
have currently completed our 2nd observing semester. To date, we have
obtained rest-frame optical spectra for 591 targeted galaxies. We have
developed a fully-automated 2D data reduction pipeline, optimized for
low S/N sources. The combination of our observational strategy and
custom reduction software leads to an improvement in S/N of up to
$25\%$ compared to standard procedures. All spectra are optimally
extracted, and all emission lines are measured using a Gaussian
fitting procedure. For galaxies without line emission, but bright
continuum emission, we measure rest-frame optical absorption lines. We
derive both continuum and line sensitivities, and show that the
theoretical expectations for the continuum emission are optimistic by
a factor of $\sim2$. For average weather conditions MOSFIRE yields a
S/N$\sim3$ per pixel within two hours in the H-band for galaxies with
a total magnitude of $H = 22$. A 5$\sigma$ detection for an emission
line within 2 hours requires a total emission line flux of $\gtrsim
1.5\times 10^{-17} \rm~erg~s^{-1}~cm^{-2}$.

With integration times of $1-2$ hours per filter, we detect multiple
emission lines for 462 out of 591 targeted galaxies. Including three
additional spectra for which we robustly identify multiple absorption
lines, we achieve a success rate of 79\%. Of the 465 confirmed
galaxies, 31 galaxies have been observed twice. Thus, the number of
unique targets with robust spectroscopic redshifts is 434. In addition
we measure robust spectroscopic redshifts for 55 galaxies that were
serendipitously detected. For 64\% of the spectroscopically confirmed
primary targets there was no robust spectroscopic redshift prior to
MOSDEF.

We derive stellar population properties for all MOSDEF galaxies by
fitting the photometric SEDs with stellar population models, while
assuming the MOSDEF redshifts. We also derive SFRs from the
combination of the \ha\ and \hb\ emission lines. The stellar masses of
our spectroscopically confirmed sample range from
$\sim~10^{9}-10^{11.5}~M_\odot$ and the SFRs range from $\sim
10^0-10^3~M_\odot~ \rm yr^{-1}$. Our spectroscopic sample exhibits a
wide variety in galaxy properties, and ranges from unobscured
star-forming galaxies, to dusty star-forming galaxies, to those
with quiescent stellar populations.

The spectroscopic success rate correlates with galaxy type, and is
highest for blue star-forming galaxies ($\sim$90\%). For red
star-forming galaxies the spectroscopic success rate is lower, and
varies from 27\% (high redshift interval) to 75\% (middle redshift
interval). The spectroscopic success rate is lowest for quiescent
galaxies, and ranges from 9\% (low redshift interval) to 50\% (middle
redshift interval). Quiescent galaxies are more challenging to confirm
as emission lines are generally faint or absent, and absorption lines
can only be detected for the brightest galaxies ($H\lesssim 22$). The
success rate is in particularly low for quiescent galaxies in the
$z\sim1.5$ sample, primarily due to poorer weather conditions, the
lack of spectra sampling the Balmer/4000\,\AA\ break regions, and the
smaller fraction of quiescent galaxies with emission lines. We have
not identified a clear cause for why the spectroscopic success rate
for red star-forming galaxies is lower than for blue star-forming
galaxies.

We compare our MOSDEF sample to the parent galaxy sample at the same
redshift from which our targets were drawn. Despite the lower success rate for
red galaxies, our prioritization by H-band magnitude ensured a
representative distribution in rest-frame $U-V$ and $V-J$ colors for
the middle and high redshift intervals. In the low redshift interval
we miss the galaxies with the reddest $U-V$ rest-frame color, due to
the lower success rate of red galaxies compared to the other redshift
intervals.

With its large sample size, its broad diversity of galaxies, its large
dynamic range in mass, SFR and redshift, and the availability of a
wealth of ancillary data in the targeted fields, the MOSDEF survey
will open up a broad range of unique science projects. Our science
objectives range from the star formation and dust properties of
distant galaxies, to the chemical enrichment history of galaxies, to
the physical properties of the ISM in the early universe, to the
accretion histories of black holes, and the structural evolution and
mass growth of galaxies over cosmic time. Early science papers, based
on data obtained during the first semester(s), focus on the relation
between gas-phase metallicity, stellar mass and SFR
\citep{RSanders2015}, the excitation properties of H\,{\sc ii} regions
\citep{AShapley2015}, dust attenuation (Reddy et al. 2014), and the
identification of AGNs \citep{ACoil2015} in $z\sim2.3$
galaxies. MOSDEF will be complemented by forefront theoretical
investigations and simulations that will help refine current models of
galaxy evolution, interstellar medium, and black hole
co-evolution. All MOSDEF data products, including the 2D and 1D
reduced spectra, spectroscopic redshifts, and value added catalogs
will be made publicly available during and upon completion of the
project\footnote{http://mosdef.astro.berkeley.edu/Home.html}.

\acknowledgements We thank the MOSFIRE instrument team for building
this powerful instrument, and for taking data for us during their
commissioning runs. M. Kriek acknowledges valuable discussion with
N. Konidaris about the reduction of MOSFIRE data and with M. Franx
regarding the noise properties of the data. We thank the referee for a
constructive report. This work would not have been possible without
the 3D-HST collaboration, who provided us the spectroscopic and
photometric catalogs used to select our targets and to derive stellar
population parameters. We are grateful to I. McLean, K. Kulas, and
G. Mace for taking observations for us in May and June 2013. We
acknowledge support from an NSF AAG collaborative grant AST-1312780,
1312547, 1312764, and 1313171, and archival grant AR-13907, provided
by NASA through a grant from the Space Telescope Science
Institute. M. Kriek acknowledges support from a Committee Faculty
Research Grant and a Hellmann Fellowship. NAR is supported by an
Alfred P. Sloan Research Fellowship. ALC acknowledges funding from NSF
CAREER grant AST-1055081. The data presented in this paper were
obtained at the W.M. Keck Observatory, which is operated as a
scientific partnership among the California Institute of Technology,
the University of California and the National Aeronautics and Space
Administration. The Observatory was made possible by the generous
financial support of the W.M. Keck Foundation. The authors wish to
recognize and acknowledge the very significant cultural role and
reverence that the summit of Mauna Kea has always had within the
indigenous Hawaiian community. We are most fortunate to have the
opportunity to conduct observations from this mountain. This work is
also based on observations made with the NASA/ESA Hubble Space
Telescope (programs 12177, 12328, 12060-12064, 12440-12445, 13056),
which is operated by the Association of Universities for Research in
Astronomy, Inc., under NASA contract NAS 5-26555.

\bibliography{mybib}

\appendix

\section{DITHER PATTERN}\label{app:dither}

\begin{figure} 
  \begin{center}           
     \includegraphics[width=0.48\textwidth,angle=0]{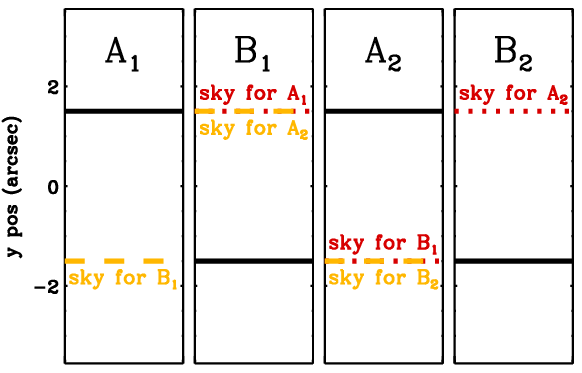}
     \includegraphics[width=0.48\textwidth,angle=0]{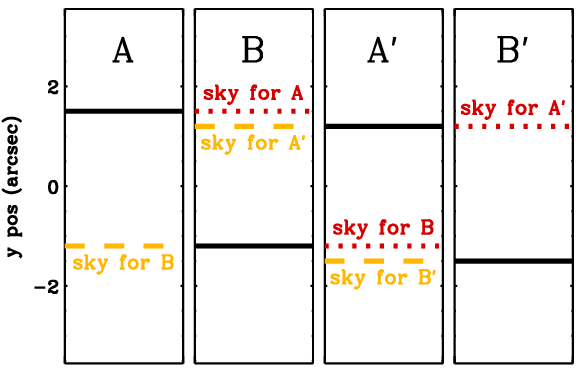}

     \caption{Illustration of the ABAB (left) and  ABA\arcmin
       B\arcmin\ (right) dither patterns. The horizontal black lines
       indicate the four dither positions for both dither
       sequences. The red dotted lines show the position of the sky
       used in the surrounding frames for the A dither positions. The
       yellow dashed lines show the position of the sky used in the
       surrounding frames for the B dither positions. This figure
       illustrate that for an ABA\arcmin B\arcmin\ dither pattern,
       different regions of the detector of frame B are used as sky
       for A and A\arcmin. For the ABAB dither pattern, the same
       regions are used as sky for the two surrounding science
       frames.}

     \label{fig:dither}
  \end{center}                            
\end{figure}

In this section we show why different dither sequences result in different S/N measurements of the reduced spectra.
For an ABBA dither sequence we use the frame before or after the exposure as sky:
\begin{equation} f_{i,s} = f_i - {f_{i \pm 1}} \end{equation}
We define the noise in a single raw image $i$ as $\sigma_i$. The noise level of the sky subtracted image then becomes
\begin{equation} \sigma_{i,s} = \sqrt{\sigma_i^2+\sigma_{i \pm 1}^2} \end{equation}
Given that the noise level of two subsequent images is approximately the same, we find:
\begin{equation} \sigma_{i,s} = \sqrt{2} \sigma_{i} \end{equation}
For an ABA\arcmin B\arcmin\ dither pattern (see Figure~\ref{fig:dither}) we can use the average of the two surrounding science frames as sky frame:
\begin{equation} f_{i,s} = f_i - \frac{f_{i-1} + f_{i+1}}{2} \end{equation}
The noise in the sky subtracted frames now becomes
\begin{equation} \sigma_{i,s} =
\sqrt{\sigma_i^2+\left(\frac{\sigma_{i-1}}{2}\right)^2 +
  \left(\frac{\sigma_{i+1}}{2}\right)^2} =
\sqrt{\frac{3}{2}}\sigma_i\end{equation}
Thus, the noise level in the sky-subtracted frames for the ABA\arcmin
B\arcmin\ pattern is $\frac{\sqrt{3/2}}{\sqrt{2}} = \sqrt{3/4}$ times lower
than the noise level for a classic ABBA dither pattern.

For an ABAB dither pattern (see Figure~\ref{fig:dither}) we can also
use the average of two surrounding sky frames to subtract from each
science frame. For an individual science frame, the noise level
decreases as well by a factor of $\sqrt{3/4}$ for this dither
pattern. However, this effect cancels out when we combine all
individual science frames to make the final spectrum, as shown below.

Consider the dither sequence $..., A_{i-1}, B_{i-1}, A_i, B_i, A_{i+1}, B_{i+1}, ...$ If we now add up three sky-subtracted science frames all at A positions we get
\begin{eqnarray}
A_{i-1,s} + A_{i,s} + A_{i+1,s} & = & A_{i-1} - \left( \frac{B_{i-2}+B_{i-1}}{2}\right) + A_{i} -  \left(\frac{B_{i-1}+B_{i}}{2}\right) +  A_{i+1} -  \left( \frac{B_{i}+B_{i+1}}{2}\right) \\
& = &  - \frac{B_{i-2}}{2} + A_{i-1} - B_{i-1} + A_{i} - B_{i} + A_{i+1}  - \frac{B_{i+1}}{2}
\end{eqnarray}
For an infinitely long sequence this becomes
\begin{equation}
A_{s} = \sum_i A_i - B_i 
\end{equation}
Thus, the ABAB dither sequence gives an approximately similar S/N level in the reduced spectrum as the ABBA dither sequence. 

However, for an ABA\arcmin B\arcmin\ dither sequence, when adding up the sky-subtracted frames $A_{i,s}$ and $A'_{i,s}$, we get the following expressing
\begin{eqnarray}
A_{i,s}(y) + A_{i,s}'(y+dy) & = & A_{i}(y) -  \left(\frac{B_{i-1}'(y)+B_{i}(y)}{2}\right) +  A'_{i}(y+dy) -  \left( \frac{B_{i}(y+dy)+B_{i}'(y+dy)}{2}\right)
\end{eqnarray}
with $dy$ the shift between $A_i$ and $A'_i$. In this
equation $B_i(y)$ and $B_i(y+dy)$ cannot simply be added, and thus
this expression cannot further be simplified. Hence, by shifting $A_i$
slightly compared to $A_{i+1}$, we use different rows of the detector
of $B_i$ as the sky frame for $A_i$ and $A'_{i}$, as also illustrated
in Figure~\ref{fig:dither}.  To obtain the maximum S/N improvement of
a factor of $\sqrt{3/4}$ in the 2D science frames, the offset between
A and A\arcmin\ needs to be at least one pixel. Otherwise $B'_{i-1}$
and $B_i$ in the above equation are not independent. For a 1D
extracted spectrum, different rows will be added together, and
thus the S/N is highest if the extraction aperture is smaller than
$dy$. For our dither sequence and average seeing conditions, this is
not the case. However, as we use an optimal extraction method, for
which most weight is given to the central rows, the dither pattern
will reduce the noise, despite the small value for $dy$.

\end{document}